\documentclass[pre,aps,amsfonts,amssymb,multicol,epsfig,graphics,referee,showpacs]{revtex4}

\usepackage{graphicx}
\usepackage[hang]{subfigure}

\newcommand \be {\begin{equation}} 
\newcommand \ee {\end{equation}}
\begin{document}

\date{\today}

\title{Statistical Physics of RNA-folding}
\author{M.\ M\"uller}
\affiliation{
Laboratoire de Physique Th\'eorique et Mod\`eles Statistiques,\\
Universit\'e Paris-Sud, b\^atiment 100, F-91405 Orsay,
France.
}

\begin{abstract}
	We discuss the physics of RNA as described by its secondary structure. We examine the static properties of a homogeneous RNA-model that includes pairing and base stacking energies as well as entropic costs for internal loops. For large enough costs the model exhibits a thermal denaturation transition which we analyze in terms of the radius of gyration. We point out an inconsistency in the standard approach to RNA secondary structure prediction for large molecules. Under an external force a second order phase transition between a globular and an extended phase takes place. A Harris-type criterion shows that sequence disorder does not affect the correlation length exponent while the other critical exponents are modified in the glass phase. However, at high temperatures, on a coarse-grained level, disordered RNA is well described by a homogeneous model. The characteristics of force-extension curves are discussed as a function of the energy parameters. We show that the force transition is always second order. A re-entrance phenomenon relevant for real disordered RNA is predicted.
 
\end{abstract}


\maketitle

\section{Introduction}  

In this paper we discuss the equilibrium statistical mechanics of RNA or single-stranded DNA as described by their secondary structure (base pairing pattern). We mainly concentrate on homogeneous polymers with uniform interactions between monomers which will be shown to capture well the physics of random disordered sequences at sufficiently high temperatures on a coarse-grained level.

We generalize the results of de Gennes' pioneering paper \cite{DeGennes68} on (homogeneous) periodic dAT-polymers (sequences $ATAT\dots$) by including entropic penalties for internal loops in order to account to some extent for self-avoidance effects. In the case where the latter are large, we predict a thermal denaturation transition that manifests itself in the scaling behavior of the radius of gyration. The scaling found in the low temperature phase is smaller than $N^{1/3}$ where $N$ is the number of monomers of the polymer. This signals an inconsistency for large $N$ since the monomer density in three-dimensional space becomes increasingly large with $N$. Excluded volume effects can therefore not be neglected in the secondary structure prediction of large molecules.

Recently diverse micromanipulation techniques have been developed that allow to monitor the response of single biomolecules, RNA or ssDNA in particular \cite{BustamanteSmith96,Maier00,BockelmannThomen02}, to an external force. These experiments have raised considerable interest in the theoretical study of force-extension characteristics of biomolecules. Within our model force-extension curves can easily be obtained upon coupling an external force to the extremities of the polymer. The molecule undergoes a thermodynamic phase transition of second order that separates the globular collapsed state from an extensive phase containing a large number of small globules \cite{MontanariMezard01}. We characterize the associated critical behavior and study in how far it is modified by the introduction of sequence randomness. Using a Harris-type criterion we argue that the correlation length remains unaffected by the disorder irrespective of temperature while other critical exponents may be modified. Numerical results indicate that at higher temperatures disordered models belong to the same universality class as the homogeneous model, while at low temperatures the collapsed phase becomes glassy and the critical behavior changes.

In a recent article \cite{ZhouZhang01} the authors claim that the introduction of base stacking energies instead of base pairing energies may change the order of the force-induced phase transition. However, they were mislead by the appearance of a sharp first order-like crossover that occurs at a higher force than the continuous phase transition described above. While the latter is almost entirely of entropic nature, reflecting the large space of energetically equivalent secondary structures, the crossover is governed by the competition between the pairing energy and the energy gained from opening and stretching it. The sharpness of the crossover results from the cooperativity due to the base stacking energy that favors long helices of stacked pairs.

\subsection{The Secondary structure of RNA}
 RNA is a linear polymer made up of four types of nucleotides, A, C, G and U. In single-stranded DNA, U is replaced by T. In solution with a sufficiently high ionic concentration to screen the charge of the phosphate backbone, the single RNA-strand has a tendency to fold back onto itself to form local double-helices of Watson-Crick base pairs (A-U and G-C) between complementary substrands of the base sequence. The entropy loss due to a bound helix is compensated for by the pairing energy due to the 2 (in A-U) or 3 (in C-G) hydrogen bonds of the base pairs and, more importantly, the stacking energy which is gained by the expulsion of water molecules between the hydrophobic parts of neighboring stacked base pairs. 

The set of all base-pairings in the RNA-molecule determines its secondary structure. The typical scale of pairing and stacking energies is considerably larger than the energy scale associated with the tertiary structure, i.e., the spatial arrangement of the RNA molecule (see \cite{BustamanteTinoco99} and references therein). This separation of energy scales is at the basis of the usual paradigm to split the RNA-folding problem into the analysis of the base pairing pattern and a subsequent determination of the tertiary structure. The set of all pairing patterns considered as secondary structures is further restricted by discarding all pairings between different loops, so-called pseudoknots. Such structures lead to knotted configurations if the helices between the loops are sufficiently long to intertwine. While knots are prevented in nature by the linear transcription process from DNA to RNA, short helices between loops can occur in principle, but they are found to constitute only a minor fraction of all base pairings. They are thus considered as elements of the tertiary interactions that can be neglected when determining the secondary structure. If we number the bases in the sequence as $i=1,\dots,N$ according to their position in the strand, the above constraint can be formalized by forbidding the coexistence of two (ordered) base pairs ($i_1$, $j_1$), ($i_2$, $j_2$) in the secondary structure with either $i_1 < i_2 < j_1 < j_2$ or $i_2 < i_1 < j_2 < j_1$. 

As we mentioned above, the separation of energy scales breaks down for large molecules, and the folding problem is complicated by the highly non-local condition that the secondary structure must have a realization in 3D. However, for intermediate degrees of polymerization $N$ the classical approach is expected to work well as is witnessed by the success of secondary structure prediction tools~\cite{ZukerSankoff84, Schuster94} that are based on the above assumptions.

In the following we will start from the usual paradigm and concentrate on the secondary structures excluding pseudoknots and other tertiary interactions. In section \ref{homfold} we discuss the statistical properties of homogeneous RNA working directly in the abstract phase space of secondary structures. This allows us to take into account systematically some excluded volume effects that reduce significantly the available configuration space of interior loops, and to gain some insight into the thermal denaturation of RNA. 
The second part of the paper is devoted to the response of RNA to an external force. In section \ref{force} the critical behavior at the force-induced opening transition is characterized and the effect of sequence disorder on the phase transition is discussed. Section \ref{extpart} deals with force-extension curves in the thermodynamic limit and its properties as a function of the energy parameters and temperature. We show that the phase transition is always of second order, but can be masked by a subsequent first order-like crossover when the cooperativity of the pairing behavior due to the stacking energy is high. 
 
\section{Folding of homogeneous RNA}
\label{homfold}

	In this section we neglect all effects due to sequence specificity. Instead we consider a RNA-model where any two bases can form a bond, their pairing affinity being independent of the bases. This exactly solvable model describes the physics of ``homogeneous'' RNA/ssDNA strands GCGCGC\dots or ATATAT\dots\cite{DeGennes68}, renormalized on the level of dimers. We will provide some evidence that random base sequences are also well described by a homogeneous model, at least at sufficiently high temperatures, if one switches to a more coarse-grained description where the monomers of the model correspond to short subunits of the single strands rather than to real bases.

\subsection{The model}

\begin{figure}
\resizebox{0.5\textwidth}{!}{
	\includegraphics{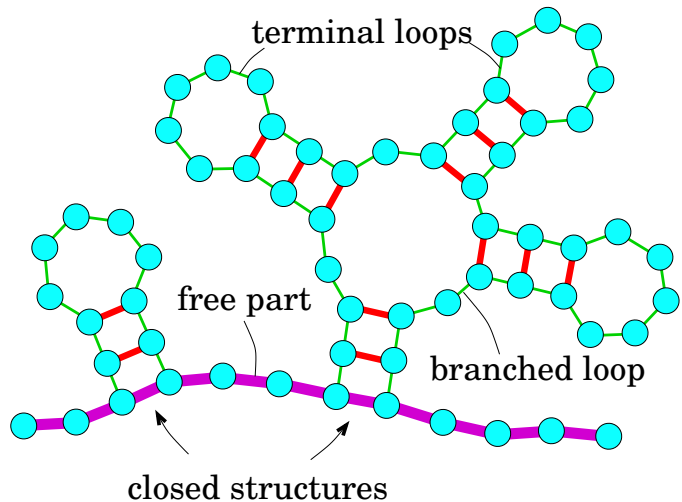}}
\caption{Elements of the secondary structure. The unpaired bases between the two closed structures constitute the free part (thick line) of the chain. The structure on the right contains a branched loop with $l=3$ unpaired bases and $m^{\prime}=m+1=4$ outgoing stems. The contour length $L$ of the loop is taken to be the number of backbone elements in the loop, i.e., $L=l+m^{\prime}$. Complementary substrands that directly fold back onto themselves form hairpins which end in terminal loops containing at least $t$ bases.}
\label{Fignotions}
\end{figure}

Following the empiric rules established by Tinoco's group \cite{Tinoco71,Tinoco73} we consider three different terms in the free energy of a given secondary structure (cf. Fig.~\ref{Fignotions} for illustration of the notions): 
Each base pair contributes the pairing free energy $f_{\rm pair}$ that we normalize with respect to the completely denatured chain where all bases are unpaired. This takes into account the mean pairing free energy (bond enthalpy and entropy cost for localization) as well as the stacking energy with the neighboring pair. In this way we count an excess stacking energy at one end of the helices which we have to compensate for by a free energy cost $-f_{\rm stack}$.

In the following these free energies will appear in the form of the temperature dependent parameters 
\be
  \eta=\exp(\beta f_{\rm stack})
\ee
and
\be
  s=\exp(-\beta f_{\rm pair}),
\ee
where $\beta$ is the inverse of the temperature $T$.
Under biological conditions (ionic strength and pH as in a living cell) $\eta\ll 1$, reflecting the importance of the stacking energy as compared to the binding energy of the hydrogen bonds. As we will see later, this is responsible for the high degree of cooperativity in the denaturation transition.

The last contribution is an entropic cost for each closed internal loop which accounts for the reduced phase space available to the loop with respect to an unconstrained string containing the same number of bases. We assume this part of the cost function to depend only on the length $L$ of the loop and the number $m^{\prime}$ of stems connected to it. The reduction of phase space gives rise to a free energy contribution of the form $\beta^{-1}\log[\phi(L;m^{\prime})]$. Specific expressions for $\phi$ will be discussed later.

When describing homogeneous RNA, we should add two further constraints on the secondary structures to be considered: First, terminal loops (the loops at the end of a hairpin) have to contain a minimal number $t$ of unpaired bases ($t=3$ from experiments) since the bending rigidity of single-stranded RNA is finite and the typical distance of a hydrogen bond $l_b$ is about 3-4 times larger than the base distance $l$ in the backbone. Secondly, stacks of less than three base pairs are unstable, and we thus require a helix to have a minimal length of $n=3$ bases.

These two conditions do not make sense for the description of real disordered base sequences on a coarse-grained level. The natural values to be taken in this case are $t=0$ and $n=1$.

\subsection{The partition function}

\begin{figure}
\resizebox{0.6\textwidth}{!}{
	\includegraphics{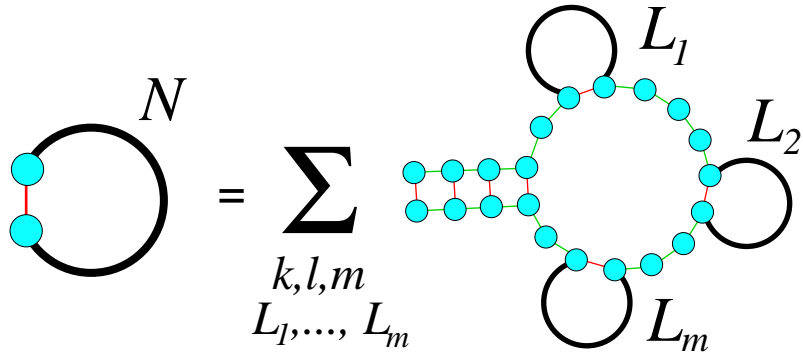}}
\caption{Schematic representation of the recursion relation for the partition function $Z^c_N$ of closed structures used to obtain Eq.~(\ref{Zrec}). The sum on the RHS is over the length $k$ of the terminal helix, the number of unpaired bases $l$ in the first loop and the number $m$ and sizes $L_i$ ($i=1,\dots m$) of closed structures attached to the loop. The shown configuration corresponds to $k=4,l=7,m=3$. The loop contributes an entropic cost $\phi(L;m')$, where $m'=m+1$ is the number of outgoing stems and $L$ is the contour length of the loop, $L=l+m'$.}
\label{Figrecursion}
\end{figure}

We denote by $Z^c_N$ the partition function of a {\em closed} RNA molecule with $N$ bases whose ends are required to form a helix. We can easily obtain a recursion relation for $Z^c_N$ (cf. Fig.~\ref{Figrecursion}): The closed secondary structure terminates in a helix containing $k\ge n$ base pairs. It is followed by a first loop which contains $l\ge 0$ unpaired bases and $m\ge 0$ closed substructures, containing $L_i\ge 2n+t$ bases, respectively. The arrangement of the free bases and substructures within the loop underlies no constraints and gives rise to a combinatoric factor. The loop contributes an entropic cost $\phi(L;m^{\prime}=m+1)$ where we take the length $L$ to be given by the number of backbone elements it contains, i.e., $L=l+m+1$.    
Finally, the sum over all configurations can be decomposed as

\begin{eqnarray}
	\label{Zrec}
	Z^c_N=\eta  \sum_{k\ge n}s^k\sum_{\{l,m\}}\sum_{L_1\ge 2n+t}\cdots\sum_{L_m\ge 2n+t}  \delta(2k+l+\sum_{i=1}^{m} L_i-N)\left( \begin{array}{c} m+l\\m\end{array}\right) \frac{1}{\phi(l+m+1;m+1)} \prod_{i=1}^m Z^c_{L_i}.
\end{eqnarray}
In the sum over $l$ (number of unpaired bases) and $m$ (number of outgoing stems) the following pairs have to be excluded: $(m=0, l<t)$ to prevent terminal loops smaller than $t$, and $(m=1,l=0)$ to avoid double counting of structures.

The structure of Eq.~(\ref{Zrec}) suggests to study the generating function $\Xi_c(\zeta)\equiv\sum_{N=2n+t}^{\infty}Z^c_N \zeta^N$ of the partition function. Taking the discrete Laplace transform we obtain
 
\be
	\label{genF}
	\Xi_c(\zeta)=\eta \frac{(s\zeta^2)^{n}}{1-s\zeta^2}\sum_{\{m, l\}}\frac{\Xi_c(\zeta)^m\zeta^l}{\phi(l+m+1;m+1)} \left( \begin{array}{c} m+l\\m\end{array}\right).
\ee

To proceed we have to make an assumption about the loop cost function $\phi(L;m)$. Neglecting loop costs altogether corresponds to putting $\phi(L;m)=1$ which yields 

\be
	\label{withoutloop}
	\Xi_c(\zeta)=\eta \frac{(s\zeta^2)^{n}}{1-s\zeta^2}\left(\frac{1}{1-\Xi_c(\zeta)-\zeta}-\Xi_c(\zeta)-\frac{1-\zeta^t}{1-\zeta}\right).
\ee
We will use this simple model in section \ref{extpart} to discuss the general shape of the force-extension characteristics. We expect it to describe well the low temperature regime where large internal loops are negligible. In order to describe denaturation we should however use a more realistic loop cost function.

\subsection{Denaturation}
\label{denaturation}
If RNA were an ideal chain without self-interaction, the entropic cost of a closed loop would just derive from the probability of a three-dimensional random walk to return to the origin, and thus $\phi(L;m)\propto L^{3/2}$ for large values of $L$. This corresponds to the case discussed in \cite{DeGennes68,MontanariMezard01} where the authors start from real space recursion relations treating the single strands as ideal chains. If one considers the loops as self-avoiding walks, forgetting about the stems that are connected to them, one is lead to use $\phi(L;m)\propto L^{3\nu_{\rm SAW}}$, with the wandering exponent $\nu_{\rm SAW}=0.588$ (in 3d) characteristic of a self-avoiding walk. (This is the form used for large interior loops in the Zuker algorithm \cite{Zukermfold}.) Clearly, this is too simple since the branches attached to the loop have a non-negligible effect on the conformational degrees of freedom of the loop and one should consider a more sophisticated form of the loop cost.

\begin{figure}
\resizebox{0.6\textwidth}{!}{\includegraphics{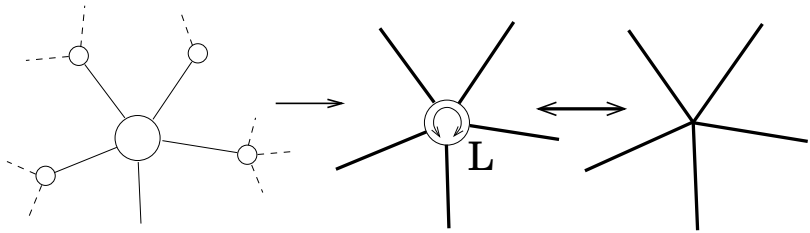}}
\caption{The effective entropy cost for internal loops of length $L$ can be obtained by idealizing the environment of the loop as $m$ outgoing rods (here $m=5$) and comparing the scaling expression for its configurational entropy with that of a star-like polymer with $m$ rays.}
\label{network}
\end{figure}

A generalization of these entropy cost functions can be obtained from the results of Duplantier and coworkers \cite{Duplantier86, Schaefer92} for the configurational entropy of a network with given topology. In order to find the scaling of the effective entropy cost of an internal loop as a function of its size we consider the secondary structure as a tree-like network of helices, linked by internal loops. Let us single out one internal loop with $m$ branches that we idealize as outgoing rods (see Fig.~\ref{network}). In the limit of loop sizes much smaller than the extension of the rods the scaling of the entropy cost for the internal loop follows from comparison of the expressions for the configurational entropy of a star-like network with $m$ rays, $\Gamma_{\rm star}(m)$, with that of a small loop with $m$ attached branches $\Gamma_{\rm loop}(L;m)$. The latter scale as $\Gamma_{\rm star}(m) \sim N^{\gamma_{\rm star}(m)-1}$ and  $\Gamma_{\rm loop}(L;m) \sim N^{\gamma_{\rm loop}(m)-1} g(L/N)$ where $N$ is the typical length scale of the attached rods (the remainder of the network), $L$ is the length of the loop and $g(x)$ is a scaling function. In the limit $L/N \rightarrow 0$ the scaling of two expressions should coincide which requires $g(x)\sim x^{\gamma_{\rm loop}(m)-\gamma_{\rm star}(m)}$. The effective loop cost function then follows as $\phi(L;m)\sim \Gamma_{\rm loop}(L;m)/\Gamma_{\rm star}(m) \sim a(m)\cdot(L)^{\nu(m)}$ where $\nu(m)= \gamma_{\rm star}(m)-\gamma_{\rm loop}(m)$. The renormalization group results from \cite{Duplantier86, Schaefer92} yield $\nu(m)= 3\nu_{\rm SAW}-m\sigma_3+\sigma_m$ where $\sigma_k$ is the exponent related to the renormalization of a vertex with $k$ legs. (See also \cite{KafriMukamel00,KafriMukamel01a,MetzlerHanke02} for a completely analogous reasoning in the closely related problems of DNA denaturation and studies of 'slip-linked' polymers). While good estimates are available via $\epsilon$- expansion for small values of $m$ \cite{Duplantier86, Schaefer92} the thermodynamics of denaturation is essentially determined by the behavior of the cost for loops with many attached stems, i.e., by $\nu(m)$ for large $m$, about which very little is known. We can proceed, however, without knowing an exact expression for $\nu(m)$. Instead we will illustrate the general condition (\ref{radofconv}) below with a discussion of the ad hoc forms $\phi(L;m)=a(m)\cdot(L)^{\nu(m)}$ with $\nu(m)=\nu^*={\rm const.}$, and $\nu(m)=\nu_0+m\nu_1$, the first one being a reasonable approximation for the case that $\nu(m)$ saturates at $\nu^*$ for large values of $\nu$, the second one assuming that each branch contributes a further entropic constraint on the loop conformations, as suggested by the term $m\sigma_3$ above. The prefactor $a(m)$ is assumed to be a moderate function of $m$ that does not grow exponentially.

The asymptotic behavior of the partition function $Z^c_N$ can be derived from the generating function $\Xi_c(\zeta)$ without performing the full inverse Laplace transform. It is given by $Z^c_N\sim\zeta_*^{-N}/N^\alpha$ where $\zeta_*$ is the smallest value of $\zeta$ at which $\Xi_c(\zeta)$ becomes non-analytic. The leading finite size corrections in the form of the pre-exponential factor $1/N^\alpha$ are determined by the nature of that non-analyticity.

There are only two possible singularities for $\Xi_c(\zeta)$: Taking successive derivatives of Eq.(\ref{genF}) one can check that all derivatives $d^k\Xi_c(\zeta)/d\zeta^k$ exist, unless the partial derivatives of both sides of Eq.~(\ref{genF}) with respect to $\Xi_c$ are equal, i.e., 

\be
	\label{touch}
	1=\eta \frac{(s\zeta_*^2)^n}{1-s\zeta_*^2}\sum_{m, l}\frac{m\Xi_c(\zeta_*)^{(m-1)}\zeta_*^l}{\phi(l+m+1;m+1)} \left( \begin{array}{c} m+l\\m\end{array}\right).
\ee
In turn this condition is sufficient to ensure a non-analyticity of $\Xi_c(\zeta)$.
Writing $\Xi_c(\zeta)=\Xi_c(\zeta_*)-\delta$ and expanding Eq.~(\ref{genF}) for small $(\zeta_*-\zeta)$ one finds $\delta^2\sim (\zeta_*-\zeta)$. Hence the singularity of $\Xi_c(\zeta)$ is approached as
\be 
	\label{singgenF}
	\Xi_c(\zeta)=\Xi_c(\zeta_*)-{\rm const.}\cdot (\zeta_*-\zeta)^{1/2}+O(\zeta_*-\zeta),
\ee
which leads to finite size corrections of the form $Z_N^c\sim\zeta_*^{-N}/N^{3/2}$. This result is central to the discussion of the critical behavior at the force-induced denaturation transition. Note that the pre-exponential factor is essentially independent of the specific choice of the model, in particular, it does not depend on the shape of the loop cost function $\phi$.
 
The second possible singularity corresponds to the double sum on the right-hand side of Eq.~(\ref{genF}) being evaluated at its radius of convergence. To analyze this situation in more detail we suppose that the loop penalty assumes the form $\phi(L;m)=a(m)L^{\nu(m)}$. According to Hadamard's formula, the radius of convergence is determined by

\be
	\label{radofconv}
	\overline{\lim}_{L\rightarrow\infty} \left[\zeta_*^L \sum_{m=0}^{L}\frac{(\Xi_c(\zeta_*)/\zeta_*)^m}{a(m+1)(L+1)^{\nu(m+1)}} \left( \begin{array}{c} L\\m\end{array}\right)\right]^{1/L}=1.
\ee

In the case where $\nu(m)$ grows at most sub-linearly in $m$, i.e., $\nu(m)/m\stackrel{m\rightarrow \infty}{\rightarrow}0$, the sum can be estimated by its saddle point, and one finds the condition $\Xi_c(\zeta_*)+\zeta_*=1$. In the same way, $\nu(m)=\nu_0+\nu_1m$ (with $\nu_1\le 1$) leads to the condition $\zeta_*=1$, but, since the sum diverges when $\zeta_*\rightarrow 1$, the singularity is ruled out in this case.

The singularity given by Eq.~(\ref{touch}) is always the smallest one at low temperatures. The system undergoes a thermodynamic phase transition when the singularity determined by Eq.~(\ref{radofconv}) crosses the first one as a function of temperature. This can only occur if the first derivative with respect to $\Xi_c$ of the double sum in Eq.~(\ref{genF}) stays finite on approaching the radius of convergence from below, that is, for $\Xi_c(\zeta_*)+\zeta_*\rightarrow 1$. This requires $\overline{\lim}_{m\rightarrow \infty}\nu(m)\ge 2$. As we will show below the corresponding phase transition is associated to thermal denaturation. In all other cases our model does not exhibit a phase transition but only a crossover whose sharpness depends both on the loop cost function and the stacking as described by the parameter $\eta$.

\subsection{Radius of gyration}

\begin{figure}
\resizebox{0.4\textwidth}{!}{\includegraphics{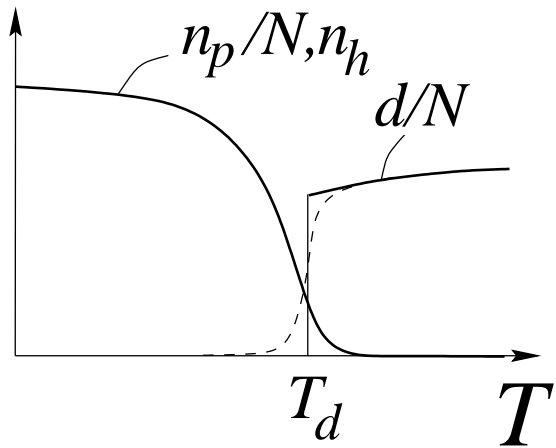}}
\caption{The denaturation transition is best described by the typical distance $d$ between bases within the secondary structure whose scaling changes from $N^{1/2}$ in the globular state at low temperatures to $N$ in the necklace-like state above $T_d$. (The dashed curve indicates $d/N$ for a molecule with finite $N$. The transition becomes sharp only in the thermodynamic limit.). In contrast, the fraction of paired bases $n_p$ or the relative number of helices $n_h/N$ only exhibit a crossover at the denaturation temperature, but never drop to zero.}
\label{FigOP}
\end{figure}

Let us now characterize the thermodynamic properties of RNA. Most observables can be obtained as appropriate derivatives of the free energy per base which is simply related to $\zeta_*$ via $f=\beta^{-1}\ln(\zeta_*)$. For example, the fraction of paired bases is given by $n_p=\partial \ln(\zeta_*)/\partial \ln(s)$ and the average number of helices is $n_h=N\partial \ln(\zeta_*)/\partial \ln(\eta)$. Evaluating these derivatives in the two possible phases one finds that $n_p$ and $n_h/N$ are both finite at all temperatures and thus do not provide a good order parameter for the phase transition (see Fig.~\ref{FigOP}). The (small) extensive number of pairings even in the high temperature phase is due to accidental pairings of bases that are close to each other within the linear RNA-strand. This result is independent of the details of the model such as the minimal number of bases in a hairpin loop, entropy costs and energy parameters.

\begin{figure}
\resizebox{0.6\textwidth}{!}{\includegraphics{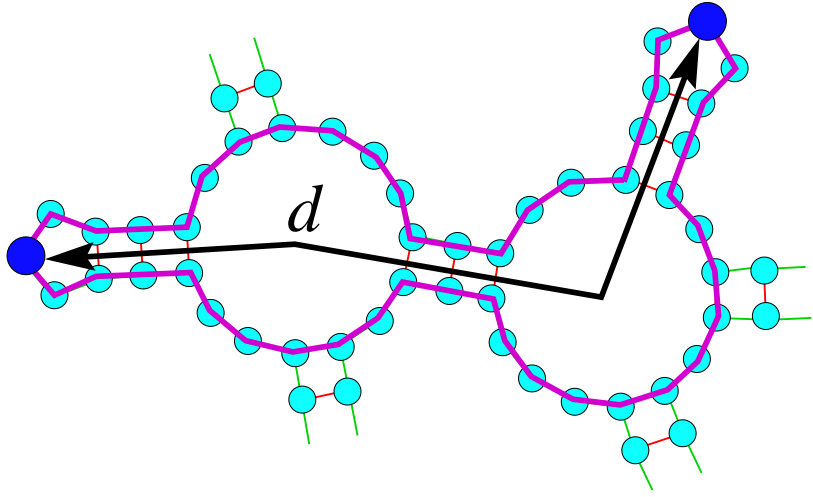}}
\caption{Definition of the distance $d$ between bases in terminal loops: The unique shortest path through helices and loops from one terminal loop to another allows us to define $d$ as the sum of the lengths of the helices and half of the contour lengths of the encountered loops, including the two terminal loops. For the bases in the figure one has $d=3/2+3+15/2+3+15/2+3+3/2$.}
\label{Figdistance}
\end{figure}

A better choice of observable is the average distance between two bases that belong to two different terminal loops of the secondary structure. This quantity is a measure for the diameter of the molecule and distinguishes the compact globular phase from the denatured loose phase. We define the distance between two bases as the length of the shortest path linking them through the secondary structure, see Fig.~\ref{Figdistance}. This path is a succession of loops and helices, its length being given as the sum of the lengths of the helices and half the contour lengths of the loops.

Alternatively, we can consider again closed secondary structures constrained to terminate in a helix. It is easy to see that the same structural information as encoded by the distance between terminal bases is captured by the average distance of terminal bases from the closing base pair $(1,N)$. The latter quantity is more convenient to compute, however.

Let us denote by $n(d;{\mathcal S})$ the number of bases in terminal loops at distance $d$ from the pair $(1,N)$. We define the Boltzmann weighted sum over secondary structures on a closed strand with $N$ bases
\be
 C_N(d)=\sum_{{\mathcal S}}Z_N^c({\mathcal S})n(d;{\mathcal S}),
\ee
for which we can write down a recursion relation in the same spirit as in Eq.~(\ref{Zrec}): The sum over all secondary structures can be decomposed into a sum over the length $k\ge n$ of the terminal helix starting at $(1,N)$, and the first loop. The latter can either be a terminal loop containing $N-2k$ unpaired bases, or there can be further closed structures connected to it. In the second case, we single out the closed structure $X$ whose terminal loops we want to consider, and denote by $m_1$, $l_1$ and $m_2$, $l_2$ the number of closed structures and unpaired bases to either side of $X$ in the loop. The $m_1+m_2$ closed structures just contribute the product of their partition functions, together with appropriate combinatorial factors for their arrangement within the loop, whereas the structure $X$ contributes the Boltzmann sum $C_{L_X}(d')$ for a distance $d'$ that is reduced with respect to $d$ by the length $k$ of the terminal helix and half of the loop contour $m_1+l_1+m_2+l_2+2$. This results in the recursion

\begin{eqnarray}
	\label{Crec}
	C_N(d)&=& \sum_{k\ge n}\eta s^k  \left[ \frac{(N-2k)\delta(d-N/2-1/2)}{\phi(N-2k+1)}
+\sum_{m_1,l_1\ge 0}\sum_{m_2,l_2\ge 0} \sum_{L_X\ge 2n+t}
\sum_{ \begin{array}{c} {\scriptstyle L_i\ge 2n+t;}\\ {\scriptstyle i=1,\dots, m_1+m_2 }\end{array} } \left( \begin{array}{c} m_1+l_1\\m_1\end{array}\right) \left( \begin{array}{c} m_2+l_2\\m_2\end{array}\right) \right. \nonumber \\ 
&&\left.  
\delta(2k+l_1+l_2+\sum_{i=1}^{m_1+m_2}L_i+L_X-N) \frac {C_{L_X}[d-k-(m_1+l_1+m_2+l_2+2)/2]}
{\phi(l_1+m_1+l_2+m_2+2)} 
\prod_{i=1}^{m_1+m_2} Z^c_{L_i}\right]. 
\end{eqnarray}

Here we used the form $\phi(L;m)\equiv\phi(L)$ for simplicity.
Passing to the Laplace transform with respect to both variables $N$ and $d$,
\begin{equation}
	\label{Laptr}
	C(\zeta,p)=\sum_{N \ge 2n+t} \sum_{d \ge n+t/2} C_N(d) \zeta^N e^{-pd},
\end{equation}
the equation is easily solved,

\begin{equation}
  \label{C}
	C(\zeta,p) =\frac{\eta(s\zeta^2e^{-p})^n e^{-p/2} g^{(t)}_\phi(\zeta e^{-p/2})}
{1-s\zeta^2e^{-p}-\eta(s\zeta^2e^{-p})^n g_{\phi}^{\prime}\left([\zeta+\Xi_c(\zeta)]e^{-p/2}\right)}.
\end{equation}

We have introduced the functions $g^{(t)}_\phi(x)=\sum_{N\ge t}N x^N/\phi(N+1)$ and $g_\phi(x)=\sum_{N\ge 1}x^N/\phi(N+1)$.

Note that $C(\zeta,p=0)$ has two possible singularities, the vanishing of the denominator and the singularity in $g_{\phi}^{\prime}$ that occurs when $\zeta+\Xi_c(\zeta)=1$. The first singularity is associated with the globular phase, the vanishing of the denominator being equivalent to condition~(\ref{touch}) in the case $\phi(L;m)=\phi(L)$. The singularity related to $g_{\phi}^{\prime}(x\rightarrow 1)$ governs the denatured phase. The denaturation transition occurs when the two singularities cross which can only happen if $g_{\phi}^{\prime}(1)$ is finite.

Let us now calculate the mean distance from bases in terminal loops to the free part. This is given by the logarithmic derivative of $C_N(p)$ with respect to $p$

\begin{equation}
 \langle  d\rangle=\partial_p C_N(p)|_{p=0}/C_N(p=0)
\end{equation}
which can be evaluated by inverse Laplace transform of $\partial_p C(\zeta,p)|_{p=0}$ and $C(\zeta,p=0)$ with respect to $\zeta$. Note that $\partial_p C_N(p)|_{p=0}$ and $C_N(p=0)$ have the same leading exponential asymptotics since their smallest singularities are the same, but their finite size corrections $\zeta_*^{-N}/N^\alpha$ differ and will determine the scaling of $\langle d\rangle$ as a function of $N$.

Let us now analyze the case of a loop cost function with asymptotics $\phi(L)\propto L^\nu$ and $\nu>2$ which implies a phase transition. In the low temperature phase, the denominator in Eq.~(\ref{C}) vanishes like $(\zeta_*-\zeta)^{1/2}$ as $\zeta$ approaches $\zeta_*$ from below (cf. Eq.~(\ref{singgenF})) which gives rise to $\alpha=1/2$ for the back-transform of $C(\zeta,p=0)$. On the other hand, the derivative $\partial C(\zeta,p)/\partial p|_{p=0}$ diverges like $(\zeta_*-\zeta)^{-1}$ which yields $\alpha=0$, whence $\langle d\rangle\sim N^{1/2}$.

The situation changes in the high temperature phase where the denominator remains finite but the generating function develops a (leading) singular part $C_{\rm sing}(\zeta,p=0)\sim(\zeta_*-\zeta)^{\nu-2}$ giving rise to $\alpha=\nu-1$. Correspondingly, the singular part of the derivative behaves as $(\zeta_*-\zeta)^{\nu-3}$, which implies $\alpha=\nu-2$, and therefore $\langle d\rangle\sim N$. 
The high temperature phase is characterized by typical distances of order $N$ from terminal bases down to the free part. This is what one expects for an essentially free, non-collapsed chain (but constrained to be paired at the ends $(1,N)$). The secondary structure is rather trivial in this case, consisting essentially of one big loop with small structures attached to it. The low temperature scaling $\langle d\rangle\sim N^{1/2}$, however, indicates the collapse to a globular state with a rich branched secondary structure.

We mention that the latter scaling can easily be derived in the absence of loop costs from the 'mountain height' representation of secondary structures \cite{BundschuhHwa01b} where the distance of terminal bases to the free part scales like the height of the mountain representation. This in turn is the typical excursion of a random walk of $N$ steps in one dimension $h$, constrained by $h>0$ and $h(1)=h(N)=0$ which is known to scale as $N^{1/2}$. There is, however, no simple equivalent of the above phase transition in the mountain height or random walk picture since the loop cost translates into an awkward non-local energy term.

The distance $d$ is a structural property of the tree-like skeleton of the secondary structure. In order to relate it to the real diameter of the molecule we assume that the helices and parts of internal loops connecting a terminal loop to the free part essentially realize a (constrained) random walk in space. If we assumed the random walk to be ideal, the radius of gyration of RNA-molecules would follow from the above findings as $R_g\sim d^{\nu_{\rm RW}}\sim N^{1/4}$ with the wandering exponent for ideal random walks $\nu_{\rm RW}=1/2$. This has already been obtained in \cite{DeGennes68}. However, the random walks should at least be considered as self-avoiding, having a larger wandering exponent $\nu_{\rm SAW}$, and correspondingly $R_g \sim N^{\nu_{\rm SAW}/2}\sim$. In addition, there are also constraints from the presence of the remainder of the molecule which have the tendency to increase the value of the wandering exponent. This effect is difficult to estimate, but we may obtain some idea about its importance by considering the two dimensional case: The wandering exponent for self-avoiding walks is known exactly as $\nu_{\rm SAW}=3/4$ \cite{Nienhuis82}. On the other hand, we need to know how the Euclidean distance $d_{\rm eu}$ between two points on a branched polymer (a coarse grained version of the secondary structure) scales with the length of the shortest path between them, the so-called chemical distance $d_{\rm ch}$. The exponent $\nu_{\rm ch}$ defined by $d_{\rm eu}\sim d_{\rm ch}^{\nu_{\rm ch}}$ is exactly known for the special case of a space-filling branched polymer where it takes the value $\nu_{\rm ch}=4/5$ \cite{Coniglio89}. It is to be expected that the exponent is slightly smaller in the case of a branched polymer of arbitrary density and thus almost equals $\nu_{\rm SAW}$. This suggests that the wandering behavior is only weakly affected by the presence of branches connected to a self-avoiding walk, and the approximation of $\nu_{\rm ch}$ by $\nu_{\rm SAW}$ is quite good.

\subsection{Discussion}

If we assume that also in the three dimensional case the presence of side branches does not increase substantially the wandering exponent from its value for the self-avoiding walk and suppose $R_g\sim N^{\nu_{\rm SAW}/2}$ we encounter a consistency problem in the thermodynamic limit: the monomer density in space diverges as $N/R_g^3\sim N^{1-3 \nu_{\rm SAW}/2}$ in the collapsed phase. This problem is common to all models considered above, irrespective of the existence of a denaturation transition. (It does not occur if the side branches have a much stronger effect than in 2 dimensions and increase the wandering exponent beyond $2/3$.) It reflects the fact that (local) entropy cost functions for interior loops are not sufficient to take into account global spatial constraints. The model customarily used in RNA prediction will thus be inconsistent for sufficiently large molecules in that it neglects excluded volume effects, deferring them to the subsequent analysis of the tertiary structure. This separation is however only justified as long as typically obtained secondary structures can easily be accommodated in space.

\begin{figure}
\resizebox{0.4\textwidth}{!}{\includegraphics{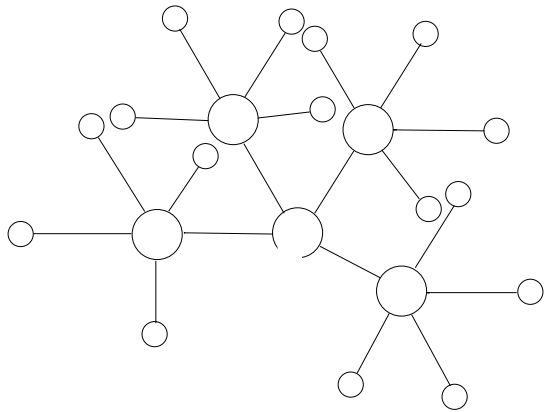}}
\caption{Schematic 'fractal' secondary structure. The large circles represent internal loops, the rods symbolize helices and the small circles hairpin loops. By choosing various complementary G-C sequences on the helices and A's in the terminal loops, e.g., it is easy to design sequences that the structure prediction algorithms predict to fold as depicted. The resulting structure is extremely dense in real space and must be discarded as an admissible folding.}
\label{fractal}
\end{figure}

In order to demonstrate that the standard RNA structure predicting programs (Zuker's mfold \cite{Zukermfold}, Vienna package \cite{Schuster94})
are indeed limited by their neglect of excluded volume effects, we have used them to determine the folding of RNA sequences that were deliberately designed to form 'fractal'
secondary structures (see Fig. \ref{fractal}). These were constructed starting from a short terminal helix that ends in a loop with two closed structures attached to it. Each of them again starts with a short helix that ends in a branched loop with two outgoing stems, and so fourth in a self-similar way. The density of such tree- or star-like structures grows exponentially with their radius. Even with a modest number of bases one can design sequences for which the structure prediction indeed yields the desired pairing pattern that can hardly be accommodated in space and should therefore be ruled out. 

While this is not a problem for small RNA molecules like transfer RNA or ribosomal RNA, the natural sizes of messenger RNA are on the order of several thousand bases which is likely in the regime where excluded volume effects play an important role for the folding and the usual structure prediction algorithms based solely on the base pairing pattern will likely fail. The inclusion of a effective loop costs to take into account the reduced available phase space might help to take into account those effects to some extent, but as we have seen they fail to cure the problem completely in the low temperature phase. 

At this point, it is worth mentioning that although neither loop costs nor the topological condition on the absence of pseudoknots are able to avoid (too) dense secondary structures, the situation would be much worse if no topological constraints were introduced in the model at all, that is, if all base pairings were allowed, irrespective of the resulting entanglement of the structure. It is rather obvious that a generic base pairing pattern obtained in such a model could not be accommodated in space: Let us consider a base pair and the two strands to which it belongs. If there is no constraint on the pairing behavior of the nearby bases within these strands, they can be paired to completely different parts of the chain which are then all forced into the same spatial region. The topological constraint forbidding pseudoknots weakens this tendency, since the substrand embraced by the given base pair is only allowed to interact with itself, which reduces largely the possibilities of spatial entanglement.   

The above observations lead to the conclusion that typical structures for large molecules are determined by a competition between favorable base pairings and the requirement that the resulting secondary structure can be accommodated in space. This will result in rather densely packed and entangled spatial arrangements of RNA that we expect to exhibit very slow dynamics and glassiness due to the inevitable spatial hindrance to pass from one favorable folded state to another. This is indeed observed in folding experiments on large ribozymes, where several misfolded states compete with the correctly folded native state \cite{Woodson00,Russell00}. Thirumalai and Woodson \cite{ThirumalaiWoodson96} propose a 'kinetic partitioning mechanism' to describe this type of slow dynamics, according to which a fraction of all molecules fold directly to the ground state while the remaining molecules remain in metastable misfolded state until they find a pathway via the transition state ensemble to the native state. Glassiness may also arise purely on the level of secondary structure \cite{Higgs96,BundschuhHwa01a,BundschuhHwa01b,PagnaniParisi00,PagnaniParisi00Reply,Hartmann01,KrzakalaMezardMuller02} where topological constraints (in the form of backbone connectivity or constraints on pseudoknots) introduce a weak frustration in the system and establish a multiplicity of metastable valleys in phase space. For small molecules this has been shown to lead to slow dynamics \cite{IsambertSiggia00}. In large molecules the jamming of the spatial arrangement of energetically favorable secondary structures probably plays an equally important role. It will be a major challenge to understand the interplay of secondary and tertiary structure and its relation with the closely related problem of protein folding.             

\section {Response of RNA to an external force}
\label{force}

\subsection{The partition function with force}

In this section we will extend our formalism to treat problems with an external force. In experiments, the RNA molecule is usually fixed on one end while the other end is manipulated by optical tweezers, magnetic beads in an inhomogeneous field, or the cantilever of an atomic force microscope. The extension of the molecule is monitored as a function of the applied force, or the position is imposed and the average force needed to maintain the position is measured. Here we will concentrate on the situation where the force is fixed while the extension is subject to thermal fluctuations.

\begin{figure}
\resizebox{0.6\textwidth}{!}{\includegraphics{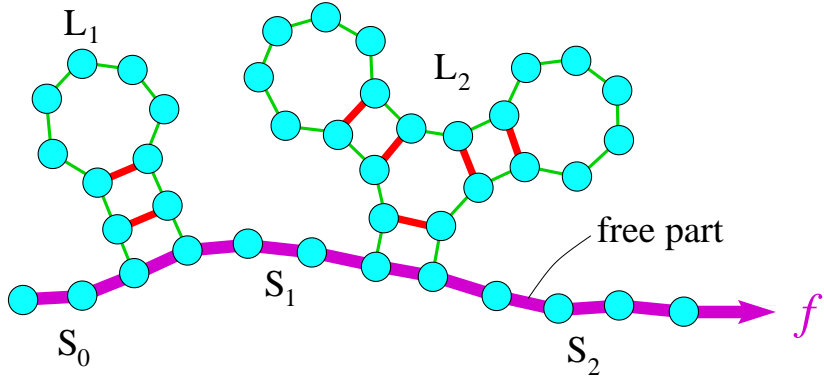}}
\caption{RNA under an external force. The force pulls on the free part of the chain that can be subdivided into single stranded portions containing $S_i$ unpaired bases ($i=0,\dots,m$). Those are separated by closed structures, containing $L_i$ bases ($i=1,\dots,m$). The free length derives from the terminating bonds of the closed structures (contributing $\zeta_b^{-1}(f)$ to the partition function) and the backbone elements in the single-stranded parts (contributing $\zeta_{\rm ss}^{-1}(f)$ in the approximation of uncorrelated monomers).}
\label{Figforce}
\end{figure}

To include the effect of an external force we have to add the term $-\vec{F}\cdot(\vec{r}_N-\vec{r}_1)$ to the energy of the system where $\vec{r}_i$ denotes the spatial position of the $i$'th base. The force only acts on the free part of the chain, see Fig.~\ref{Figforce} and we can rewrite the additional term as

\be
	-\vec{F}\cdot(\vec{r}_N-\vec{r}_1)=-\sum_{i=1}^{l_b-1}\vec{F}\cdot(\vec{r}_{b(i+1)}-\vec{r}_{b(i)}),
\ee 
where $\{b(i)\}$ is the ordered list of all bases in the free part, and $b(l_b)=N$. There are two types of contributions to the sum: Terms with $b(i+1)=b(i)+1$ correspond to successive bases in the backbone of the RNA while terms with $b(i+1)>b(i)+1$ correspond to the paired bases terminating the closed structure between bases $b(i)$ and $b(i+1)$. For simplicity we consider both the distance $l$ of bases within the backbone and the distance $l_b$ of covalently bonded bases as fixed. We treat the terminating hydrogen bonds of closed structures as free joints that are inserted between single strands of unpaired bases. They contribute a factor $1/\zeta_b(f)=\sinh(\beta f l_b)/(\beta f l_b)$ to the partition function. For the single strands in between we will restrict ourselves to the simple model of a freely jointed chain. If one wants to fit to experimental data \cite{MontanariMezard01}, one should however consider a more realistic description involving correlations of the monomers on the scale of the persistence length ($l_p\approx 3l$). At high forces, bond elasticities should also be included. Both modifications can straightforwardly be taken into account in the formalism below.

The partition function $Z_N^f(f)$ including the external force can easily be obtained once the partition function $Z_N^c(f)$ of closed structures is known. We may decompose the sum over all structures into a sum over the number $m$ and sizes $L_i$ ($i=1,\dots m$) of closed structures in the free part, and the lengths $S_i$, ($i=0,\dots m$) of the single-stranded segments linking them:

\be
	\label{recZf}
	Z^f_N(f)=\sum_{m\ge0}\sum_{S_0,\dots,S_m\ge0}\sum_{L_1,\dots,L_m\ge0} \delta(S_0+\sum_{i=1}^m (L_i+S_i)-N) Z_{S_0}^{\rm ss}\prod_{i=1}^{m-1}\left[\frac{Z^c_{L_i}Z_{S_i+1}^{\rm ss}(f)}{\zeta_b(f)}\right]\frac{Z^c_{L_m}Z_{S_m}^{\rm ss}(f)}{\zeta_b(f)}.
\ee
Here we have introduced the partition function $Z^{\rm ss}_N(f)$ of a single-stranded segment with $N$ backbone elements under the force $f$. The corresponding generating function follows from a discrete Laplace transform as

\begin{eqnarray}
	\label{genFf}
	\Xi_f(\zeta;f)\equiv\sum_{N}Z^f_N(f)\zeta^N
	= \frac{\Xi_{\rm ss}^2(\zeta;f)\zeta}{\left(\Xi_{\rm ss}(\zeta;f)-1\right)\left(1-\frac{\Xi_c(\zeta)}{\zeta_b(f)}\frac{\Xi_{\rm ss}(\zeta;f)-1}{\zeta}\right)}.\nonumber
\end{eqnarray}

The partition function is again found by inverting the Laplace transform, in particular, the free energy derives from the logarithm of the smallest singularity of $\Xi_f(\zeta;f)$. Apart from the singularities of $\Xi_c(\zeta)$ that we discussed in section \ref{denaturation}, $\Xi_f$ also has a pole singularity $\zeta(f)$ when the denominator in Eq.~(\ref{genFf}) vanishes:

\be
	\label{zetaf} 
	1=\frac{\Xi_c(\zeta(f))}{\zeta_b(f)}\frac{\Xi_{\rm ss}(\zeta(f);f)-1}{\zeta(f)}.
\ee

Let us now fix the temperature. The singularity deriving from $\Xi_c$ then takes the force-independent value $\zeta_*=\zeta_*(T)$. A phase transition occurs at the critical force $f_c(T)$ where $\zeta(f)$ crosses $\zeta_*(T)$. For larger forces, the force-extension characteristics follow from $\langle L(f)/N\rangle\stackrel{N\rightarrow\infty}{=}-\beta^{-1}\partial \ln[\zeta(f)]/\partial f$ in the thermodynamic limit. Here, $L(f)$ denotes the projection of the end to end distance of the molecule onto the direction of the force.    

In the following we are interested in two aspects of the force-extension curve. First, we will examine in detail the critical behavior around $f_c$. Later, we address the temperature dependence of $f_c(T)$ and an associated re-entrance phenomenon slightly below denaturation. Finally, in section \ref{extpart} we will discuss in detail the dependence of the force-extension characteristics on the parameters $s$ and $\eta$.

\subsection{The critical behavior around $f_c$}

In the following we treat the single stranded parts as freely jointed chains, whose generating function is given by 
\begin{equation}
	\label{genfss}
	\Xi_{\rm ss}(\zeta;f)=1/(1-\zeta/\zeta_{\rm ss}(f)),
\end{equation}

where $\zeta_{\rm ss}(f)=\sinh(\beta fl)/(\beta fl)$. Furthermore, we only consider temperatures below denaturation. 

In the vicinity of the (finite) critical force we may restrict ourselves to the relevant singularity structure of $\Xi(\zeta;f)$ and expand the denominator in Eq.~(\ref{genFf}) to lowest non-trivial order around $f_c$ and $\zeta_*$. Using Eq.~(\ref{singgenF}) we find

\be
	\label{apprXif}
	\Xi_f(\zeta;f)\approx \frac{B}{(1-\zeta/\zeta_*)^{1/2}-A (f-f_c)},
\ee
or, on substituting $\zeta\equiv e^{-s}$ and $\zeta_*\equiv e^{-s_*}$,
\be
	\label{apprXifs}
	\Xi_f(s;f)\approx \frac{B}{(s-s_*)^{1/2}-A (f-f_c)}.
\ee

$A$ and $B$ are slowly varying functions of $f$ and $\zeta$ that we replace by their values at the critical point, $A\equiv A(f_c,\zeta_*)$ and $B\equiv B(f_c,\zeta_*)$. The (continuous) inverse Laplace transform of Eq.~(\ref{apprXifs}) is explicitly known, and we obtain the partition function in the transition region as

\be
	\label{apprZf}
	Z^{f}_N(f)=\frac{ B e^{s_*N}}{\sqrt{\pi N}} \psi(A(f-f_c)N^{1/2}),
\ee
where $\psi(x)=1+\sqrt{\pi}x\exp(x^2){\rm erfc}(-x)$. The force-extension characteristics follow immediately as $L(f)=AN^{1/2}(\ln \psi)'[A(f-f_c)N^{1/2}]$. In the asymptotic regimes of the scaling variable $x=A(f-f_c)N^{1/2}$ one obtains the expansions

\be
	\label{fl}
	\beta L(f) \approx \left\{ 
	\begin{array}[c]{ll} 2A^2(f-f_c)N\left[1+\frac{1}{2x^2}+O\left(\frac{e^{-x^2}}{x^3}\right)\right] & x\gg 1,\\
	AN^{1/2}\sqrt{\pi}\left[1+\frac{4-\pi}{\sqrt{\pi}}x+O({x^2}))\right] & |x|\ll 1, \\
	\frac{2}{f_c-f}\left[1-\frac{3}{2x^2}+O(\frac{1}{x^4})\right] & x\ll -1.
	\end{array}	
	\right.
\ee

Sufficiently above the critical force ($x\gg 1$), the extension grows linearly with $f-f_c$ and scales as the system size. The chain organizes in a kind of necklace: the number of closed structures in the free chain is proportional to $N$, their average size being finite. In the low force regime ($x\ll 1$) the chain is collapsed, but its extension diverges as $1/(f_c-f)$ upon approaching the critical point.

There are two critical exponents characterizing this phase transition:
At the critical force ($x\ll 1$) the extension obeys a power law $L\sim N^{\delta}$ with $\delta=1/2$. The second exponent is related to the characteristic length scale in the problem, $N_c\propto (f-f_c)^{-\nu}$ where $\nu=2$, as one can read off from the form of the scaling variable. Below we will see that $N_c$ can be understood as a correlation length.

\subsection{Correlations and length scales at $f\ge f_c$}

Slightly above the critical force, the typical number of bases in a closed structure is given by $l_{\rm typ}=N/n_{\rm cs}(f)$ where $n_{\rm cs}(f)$ is the number of closed structures in the free part. It has the same critical behavior as the extension, i.e., $n_{\rm cs}(f\ge f_c)\propto N(f-f_c)$, and thus, $l_{\rm typ} \propto 1/(f-f_c)$. This is surprising since the characteristic length scale $N_c\propto (f-f_c)^{-2}$ diverges much faster.

To understand the meaning of $N_c$ let us introduce the indicator function $\eta_i$ which equals 1 if base $i$ belongs to the free part, and 0 otherwise.
The correlation function $\langle \eta_i\eta_j\rangle$ is simply obtained as the ratio between the partition function with bases $i$ and $j$ constrained to be free, and the total partition function,

\be 
	\langle \eta_i\eta_j\rangle=\frac{Z^f_{i-1}Z^f_{j-1-(i+1)}Z^f_{N-(j+1)}}{Z^f_N}.
\ee

Using Eq.~(\ref{apprZf}) we obtain the connected correlation function as
\be 
	\label{conncorr}
	\frac{\langle \eta_i\eta_j\rangle-\langle \eta_i\rangle\langle \eta_j\rangle}{\langle \eta_i\rangle\langle \eta_j\rangle}=\frac{Z^f_{j-i-2}Z^f_N}{Z^f_{N-(i+1)}Z^f_{j-1}}-1\approx \frac{e^{-(j-i-2)[A(f-f_c)]^2}}{4\sqrt{\pi}(j-i-2)^{3/2}[A(f-f_c)]^3},
\ee
where the last approximation is valid in the scaling regime above the critical force for $i,N-j,N\gg (j-i)\gg [A(f-f_c)]^{-2}$. The quantity $N_c=[A(f-f_c)]^{-2}$ clearly appears as the correlation length beyond which the pairing behavior becomes essentially independent. To see why the correlation length is much larger than typical closed structures, let us look at the probability distribution of the sizes of the latter.

Suppose that a closed structure starts at base $i$. The probability that it is paired to the base $j=i+l+1$ is given by
\begin{eqnarray}
	\label{Prob(l)}
	P(l)&=&\frac{Z^f_{i-1}Z^c_{l}Z^f_{N-(i+l+2)}}{\sum_{l'>0}Z^f_{i-1}Z^c_{l'}Z^f_{N-(i+l'+2)}} \propto \frac{\exp(-lA^2(f-f_c)^2)}{l^{3/2}}
\end{eqnarray}
from which we recover the expectation value for the structure size $l_{\rm typ}=\langle l\rangle=\sum_l lP(l)\propto 1/(f-f_c)$. On the other hand, we can calculate the fraction $\chi(l_*)$ of bases that belong to closed structures of size at least $l_*$:

\be 
	\frac{\sum_{l= l_*}^\infty lP(l)}{\sum_{l= 1}^\infty lP(l)}\approx \frac{\int_{l_*[A(f-f_c)]^2}^\infty x^{-1/2}\exp(-x)\, dx}{\int_{0}^\infty x^{-1/2}\exp(-x)\, dx}={\rm erfc}[A(f-f_c)l^{1/2}_*].
\ee
A finite fraction of all bases thus belong to structures of size $O((f-f_c)^{-2})$ which sets the scale of the correlation length $N_c$. The vast majority of closed structures is much smaller, however.

\subsection{The critical behavior with sequence disorder}
After having understood the critical behavior in the homogeneous case, it is natural to ask whether disorder in the form of sequence inhomogeneities and varying pairing affinities between the bases is a relevant perturbation for the force-induced phase transition. In \cite{LubenskyNelson00,LubenskyNelson02} the authors studied the force-induced unzipping of DNA and found the presence of disorder to significantly alter the critical behavior with respect to that of a homogeneous double-strand. In RNA, the disorder effects are less pronounced since the two opening transitions are not really of the same nature. In DNA, essentially all base pairs are broken up at the transition and the double-strand becomes denatured. The force always acts only on the single base pair closing the yet unzipped double helix. In RNA, however, the transition occurs at a point where the entropy of large secondary structures and the free energy gain from the extension of the chain compete in a quite subtle manner, the base-pairing energies playing a less important role at the critical force. Furthermore, already in the critical region the force acts in parallel on a large number of globular structures aligned along the free part of the chain, which averages out the effect of disorder to some extent \cite{GerlandBundschuh01,MullerKrzakalaMezard02}.

In the case of RNA will be interested in the low-temperature regime where we can simplify the model by neglecting the loop cost function, i.e., putting $\phi\equiv 1$. Furthermore, we replace the pairing and stacking free energies by simple (temperature-independent) pairing energies $e_{ij}$ between the bases $i$ and $j$. This does not change the critical behavior at the force-transition in the homogeneous case ($e_{ij}\equiv e$), and we checked numerically that a disordered model with pure stacking energies leads qualitatively to the same results as the pairing model.

As in earlier work \cite{BundschuhHwa01b, KrzakalaMezardMuller02, MarinariPagnaniRicci02} we consider different types of disordered models. The most natural one starts from RNA made of the four base species $b_i\in \{A,C,G,U\}$. The pairing energies will then depend on the sequence via $e_{ij}=E(b_i,b_j)$ where $E$ is a symmetric $4\times 4$ matrix. We used the simple matrix $E(C,G)=E(G,C)=-3$, $E(A,U)=E(U,A)=-2$ (Watson-Crick pairs) and $E(G,U)=E(U,G)=-1$ (wobble pairs), and $E=+\infty$ for all other pairs. Alternatively we considered more abstract random coupling models where the $e_{ij}$ are independent variables taken from a distribution $P(e)$. In the following we focus on the two cases where $P(e)$ is Gaussian or has power law tails decaying like $|e|^{-\alpha}$, respectively, both being centered on a negative value.



\begin{figure}
\resizebox{0.5\textwidth}{!}{\includegraphics{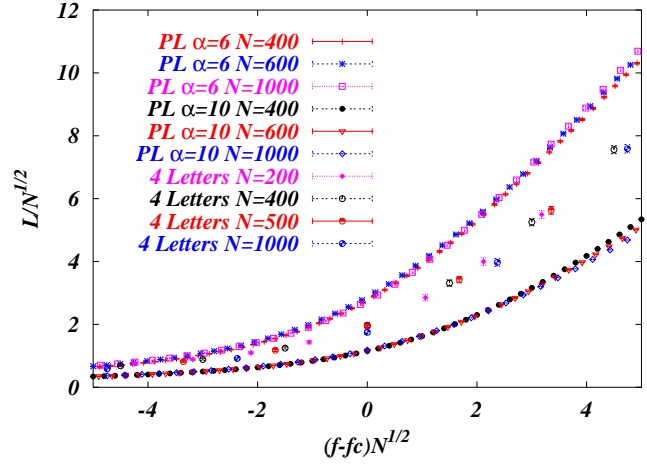}}
\caption{Scaling plot of force-extension curves for disordered models (power law distributions (PL) with $\alpha=6$ and $\alpha=10$, and the 4 letters model) at high temperature ($T=0.6$). For better visibility, the extension of the 4 letters model has been multiplied by $1.5$.}
\label{scalinghighT}
\end{figure}

\begin{figure}
\resizebox{0.5\textwidth}{!}{\includegraphics{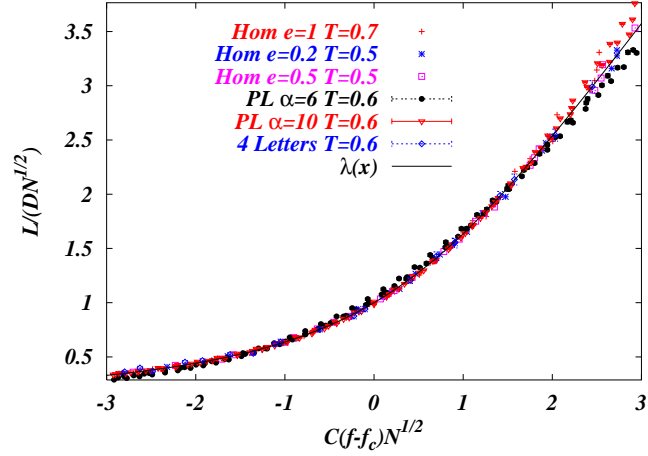}}
\caption{Force-extension curves from Fig.~\ref{scalinghighT} and for homogeneous models at different temperatures $T$ and pairing energies $e$. All curves superpose with the analytical prediction (\ref{fl}) $L(f)=DN^{1/2}\lambda(C(f-f_c)N^{1/2})$ upon rescaling with model-dependent factors $C, D$. The error bars are smaller than the symbol sizes.}
\label{universality}
\end{figure}

The numerical evaluation of force-extension characteristics for these types of models is straightforward using the $O(N^3)$ recursion relation as introduced in \cite{Waterman78,McCaskill90} to compute the partition function $Z^f_N$ exactly for a given realization of the disorder (see Refs. \cite{GerlandBundschuh01} for a related investigation, and \cite{MullerKrzakalaMezard02} for a more thorough discussion of the effects of disorder and the low temperature behavior). In Fig.~\ref{scalinghighT} we show scaling plots of the disorder-averaged force-extension characteristics for several disordered models at temperatures well above the glass transition temperature \cite{KrzakalaMezardMuller02,BundschuhHwa01b}. The data collapse in the critical regime was obtained optimizing the critical force in the scaling ansatz $L(f)=N^{1/2}\lambda((f-f_c)N^{1/2})$ supposing that the critical exponents are the same as in the homogeneous case ($\delta=1/\nu=1/2$). The scaling works well for the model with four letters and for the random coupling models with a Gaussian probability distribution, or with $P(e)\sim e^{-\alpha}$ and $\alpha>4$. However, the data obtained for distributions $P(e)\sim e^{-\alpha}$ with $\alpha\le 4$ (not shown) cannot be collapsed satisfactorily even when allowing for different critical exponents. This is due to the dominance of some rare but very strong couplings as will be explained in the next subsection.

By rescaling the axes of the plots with model-dependent metric factors $C$ and $D$, $L(f)=DN^{1/2}\lambda(C(f-f_c)N^{1/2})$, one can perfectly superpose the scaling functions for the different models with that of the homogeneous model, as shown in Fig.~\ref{universality}. This indicates that, above the glass transition temperature, disorder is irrelevant for the force-induced phase transition in the sense that all sufficiently short-ranged disordered models fall into the same universality class as homogeneous RNA. The latter suggests that the behavior of disordered RNA at high temperatures is well captured by a coarse-grained homogeneous description with renormalized parameters. The effect of disorder is washed out by thermal fluctuations which allow for a large number of secondary structures to be explored so that the large entropy of secondary structures (with approximately the same number of bonds) dominates the physics. 

\begin{figure}
\resizebox{0.7\textwidth}{!}{\includegraphics{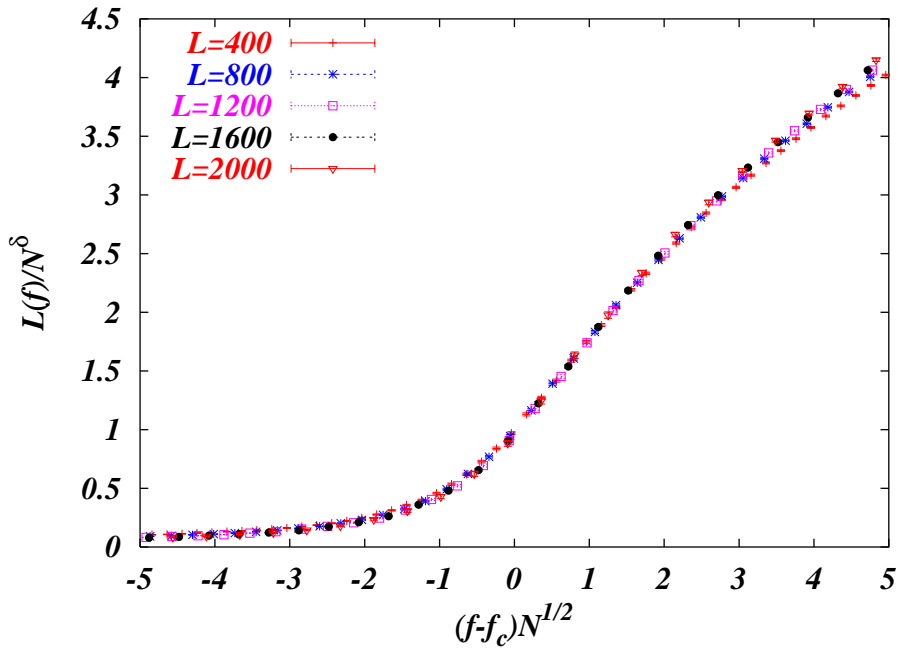}}
\caption{Scaling plot of force-extension curves for the Gaussian model at T=0. The critical exponent is modified by the disorder to $\delta\approx 0.7$.}
\label{fl0curves}
\end{figure}

The situation is different at low temperatures where the molecule is restricted to a small number of favorable foldings. The disorder-averaged force-extension curves at zero temperature are shown in Fig.~\ref{fl0curves} where the data collapse has been achieved with the general ansatz $L(f)=N^{\delta}\lambda((f-f_c)N^{1/\nu})$, optimizing for $f_c$, $\delta$ and $\nu$. As we will discuss in the next paragraph, the correlation length exponent $\nu=2$ stays unchanged with respect to the homogeneous case. However, the exponent $\delta$ is modified ($\delta=0.7$) \cite{MullerKrzakalaMezard02}. 

\subsection{Harris-type criterion for the relevance of disorder}            
The relevance of disorder for a phase transition can often be judged by applying Harris' criterion according to which disorder is relevant if the specific heat exponent $\alpha=d\nu-2$ is positive. Plugging in $d=1$, since the sequence of bases is one-dimensional, and using the correlation length exponent $\nu=2$, we are lead to conclude that disorder is marginal for the force transition. 

In order to derive this result, we start from the homogeneous model where the pairing behavior is correlated up to the scale $\xi\sim |f-f_c|^{-2}$. The introduction of disorder will locally modify the value of the critical force, $f_c\rightarrow f_c+\Delta f_c$ whereby this only makes sense as long as one considers substrands that are large compared to the bare correlation length. To estimate the typical fluctuations $\Delta f_c(\xi)$ for regions of size $\xi$ we observe that the opening transition is mainly of entropic nature (see section \ref{extpart}) and results from the competition between the gain in stretching energy and the decrease of the number of possible secondary structures when the chain changes from a globular state to a necklace with a larger number of closed substructures.

The effect of the pairing energies comes only as a perturbation. It is thus reasonable to expect that $\Delta f_c(\xi)$ scales like the fluctuations of the average binding energy per base for secondary structures that are restricted to a substrand of length $\xi$. This implies $\Delta f_c(\xi) \sim \xi^{-1/2}$.

Locally, the correlation length is modified according to
\be
	\label{corrloc}
	\xi\sim |f-f_c-\Delta f_c(\xi)|^{-2}.
\ee
The scaling $\Delta f_c(\xi)\sim \xi^{-1/2}\sim (f-f_c)$ corresponds just to the limiting case for which the critical force is still uniquely defined and the exponent $\nu=2$ remains unchanged. This reflects the marginality as predicted by the vanishing of $d\nu-2$.

The above considerations are wrong if the disorder distribution has large tails in which case very large though rare couplings may dominate the secondary structure pattern. For the random coupling model with power law tails $P(e)\sim |e|^{-\alpha}$ we can easily find a lower bound on $\alpha$ below which disorder will significantly alter the behavior of the model, rendering it even non-self-averaging. The energy fluctuations will scale like the energy of rare favorable secondary structures. We may estimate the latter by considering a ``greedy'' algorithm that constructs a secondary structure by choosing iteratively the base pair with the most negative energy available while respecting the topological constraints imposed by the pairs already chosen. There are $N(N-1)/2$ pairing energies available for the first step, and the best among them will scale as $N^{2/\alpha}$. There will be of the order of $\log(N)$ further choices that lead to comparable energies, while in later stages the pairing energies will be significantly smaller. We thus expect the disorder induced energy (fluctuations) to scale at least as $\Delta E(N)\sim N^{2/\alpha} \log(N)$, or $\Delta f_c(N)\sim N^{2/\alpha-1} \log(N)$. Thus, the fluctuations dominate for $\alpha\le 4$. Indeed, we did not succeed in collapsing the numerical data for $\alpha=3,4$.

\subsection{Why is the force induced transition of the second order?}
It is rather unusual to find a continuous phase transition in force-extension experiments. The closely related globule-coil transition in polymers is of the first order which is a consequence of the large finite size corrections to the free energy of the globular phase of a chain with $N$ elements, $F_{\rm gl}(N)= f_{\rm gl}N+aN^{2/3}$. The term $aN^{2/3}$ takes into account solvent effects at the surface. Such corrections are essentially absent in the (extensive) random coil phase since all monomers are more or less in a similar environment; but the free energy depends on the external force $f$ since the structure is extensible, $F_{\rm coil}(N)=Nf_{\rm coil}(f)$. At the force where the extensive parts of the free energy become equal, $ f_{\rm gl}=f_{\rm coil}(f)$, a discontinuous transition from the globular to the random coil phase takes place \cite{GeisslerShakhnovich02a, GeisslerShakhnovich02b}. Mathematically, the first order nature of the force transition is reflected by the an essential rather than algebraic singularity in the Laplace transform of the partition function in the globular phase. 

In our model for RNA a surface term in the globular phase is absent since solvation energies and surface effects are part of the tertiary interactions that are far less important than the base pairing (at least for small and intermediate sizes). The finite size corrections of the free energy in the globular phase are only of order $\log(N)$. Thus, a subdivision of the chain into a necklace of globules is less costly than in the presence of surface effects. At the critical force, this leads to a continuous crossover from a single large globule to a necklace containing an extensive number of smaller globules which takes place over a force window decreasing as $N^{-1/2}$. In a two-dimensional homogeneous model of the globule-coil transition \cite{Marenduzzo02b}, the authors found the force-transition to be continuous, too, which can be traced back to the absence of surface energies that grow polynomially with the system size.

It is worth mentioning that the thermodynamic phase transition would be absent in our model if the secondary structures were not allowed to contain multi-branched loops, but were limited to single hairpins (with possible alignment gaps). Instead, there would only be an opening crossover. Although the continuous phase transition is an otherwise robust feature of all models irrespective of the details of the pairing and stacking rules, it critically depends on the topological constraints.

\section{Discussion of force-extension curves in the thermodynamic limit}
\label{extpart}

In a recent paper~\cite{ZhouZhang01} the authors claim that the inclusion of large stacking energies in the model renders the force transition first order in contrast to the second order transition found in a model with only pairing energies. This has resulted from an erroneous analysis of a system of equations for generating functions that are real space analogs of our Eqs.~(\ref{genFf}) and (\ref{genF}). The authors were mislead by a sharp force-induced denaturation crossover that masks the true thermodynamic transition at a smaller force where the extension begins to grow only very slowly as a function of force.

Before we discuss the general properties of force-extension curves, let us recall the parameters entering the model: Helices are required to contain at least $n$ base pairs, and terminal loops closing a hairpin consist of at least $t$ unpaired bases. The distance between monomers in the backbone is denoted by $l$, while $l_b$ is the average distance between covalently bound bases. If we describe homogeneous RNA, we should use the empirically determined values  $l_b\approx 4l$, $n=3$ for the minimal helix length and $t=3$ for the minimal hairpin loop. However, if the model is used to describe disordered RNA on a coarse-grained level, natural values are $l_b\ll l$, $n=1$ and $t=0$.

The base pair interactions are described by two parameters, $s(T)$ and $\eta(T)$, accounting for the pairing and stacking energy per base pair, and the cost for the initiation of a helix, respectively. As mentioned earlier, under physiological salt conditions the cooperativity parameter $\eta(T)$ is very small and thus favors the formation of long helices. The parameter $s(T)=\exp(-\beta f_{\rm pair})$ is large at sufficiently low temperatures but approaches $s\approx 1$ in the denaturation regime. In the following discussion we will consider $\eta(T)$ to be small throughout and $s(T)$ to be large at low temperatures, while approaching $1$ around the denaturation temperature $T_d$. We distinguish the three temperature regimes, dropping the explicit temperature dependence of $s$ and $\eta$: $s\gg 1$, $1\gg s-1\gg \eta^{1/3}$ and $0<s-1\ll \eta^{1/3}$. The case $s<1$ corresponds to denatured RNA which is not of interest for force-extension studies.

For analytical simplicity we use the model without loop cost function, $\phi\equiv 1$, see Eq.~(\ref{withoutloop}). This is expected to be justified in the low temperature regime $s\gg1$, as well as at high forces, while the results for $s\approx 1$ at low force have to be taken with some care.

For the details of the calculations the reader should refer to the appendix.

\subsection{Critical force and re-entrance}

In order to discuss the thermodynamic limit of force-extension curves, we need the free energy per base, $\phi(f)$, as a function of the force. It follows via $\zeta(f)\equiv\exp[-\beta \phi(f)]$ from Eq.~(\ref{zetaf}) that we rewrite as 
\be
	\label{zetaf2}
	\zeta_{\rm ss}(f)=\zeta(f)+\frac{\Xi_c(\zeta(f))}{\zeta_b(f)}.
\ee
Here we treat the single strands linking the closed structures as freely jointed chains, whose free energy per base is related to $\zeta_{\rm ss}(f)$ via $\zeta_{\rm ss}(f)\equiv\exp[-\beta \phi_{\rm ss}(f)]$. The free energy per base in the globular phase, $\phi_*$, is determined from the singularity $\zeta_*=\exp(-\beta \phi_*)$ of $\Xi_c$.

The chain begins to open when both free energies are equal, i.e., when $\phi_*=\phi(f)$, and the critical force is determined by the equation $\zeta_*=\zeta(f_c)$.

In the low temperature regime ($s\gg 1$), the critical force depends on the parameter $t$ for the minimal length of terminal loops. For $t\ge 1$ we find 

\be
	\label{fcexplicit}
	f_c(T)\approx\frac{t}{4(l_b+l)}|f_{\rm pair}(T)|.
\ee 	
The dependence on $t$ is due to the fact that each hairpin terminates in a loop with at least $t$ unpaired bases. The corresponding loss in energy is very important at low temperatures and limits the thermally accessible phase space to rather elongated hairpinned structures with few branchings. This manifests itself in the critical force being proportional to the pairing free energy, almost as in the unzipping of DNA.
  
For $t=0$ the situation is different since there is no energy cost associated to a hairpin, and therefore the available phase space of secondary structures is still large, even at low temperatures. The phase transition is governed by the competition between the force, trying to increase the number of closed structures in the free part, and the entropy that favors one big closed structure. The equation for the critical force reduces to 
\be
	\label{fct=0}
	2\zeta_{\rm ss}(f_c)\zeta_{b}(f_c)\approx 1
\ee
and thus $f_c(T)\propto T/l$, almost independently of $s$. This reflects the purely entropic origin of the critical force sufficiently below denaturation .

Clearly, $t=0$ corresponds to an unphysical situation if the monomers in our model are interpreted as nucleotides. However, if we regard the homogeneous model as a coarse-grained description of a disordered base sequence, the monomers in the model stand for short substrands with an average affinity to pair with other substrands. The frustration in the secondary structure of disordered RNA necessarily leads to gaps in the base pairing that are usually larger than the minimal length of terminal loops. It is therefore unnecessary to impose a constraint on the terminal loops, i.e., we may safely put $t=0$ in this case. At the same time, the minimal length $n$ of helices and the length parameters $l$ and $l_b$ have to renormalized appropriately, as we indicated earlier.

In the denaturation regime, $s\approx 1$, the critical force becomes small. Independently of $t$, it decreases as

\be
	\label{fchighT}
	f_c(T)=\left\{ \begin{array}[c]{ll}
	 O((s-1)^{1/2}) & 1\gg s-1\gg \eta^{1/3}\\
	 O(\eta^{1/6}) & s-1\ll\eta
	\end{array}	\right. 
\ee
on approaching denaturation.

For $t=0$ an interesting re-entrance phenomenon occurs: We have seen that in this case $f_c(T)$ is an increasing function of temperature at sufficiently low temperatures, which is due to the entropic nature of the critical point. The large value of the binding parameter $s$ merely forces the dominant secondary structure to have all bases paired, but does not influence the critical behavior otherwise. This picture does, however, not apply near denaturation where the base pairs are only loosely bound. Rather, Eq.~(\ref{fchighT}) shows that $f_c$ decreases essentially to zero in the denaturation regime. This is a consequence of the vanishing of the pair binding free energy ($\propto s-1$) which competes with the free energy gained from stretching. The latter grows as $f^2$ at low forces, and thus, the phase transition takes place on a force scale of the order of $(s-1)^{1/2}$ which vanishes at the denaturation transition.

\begin{figure}
\resizebox{0.5\textwidth}{!}{
	\includegraphics{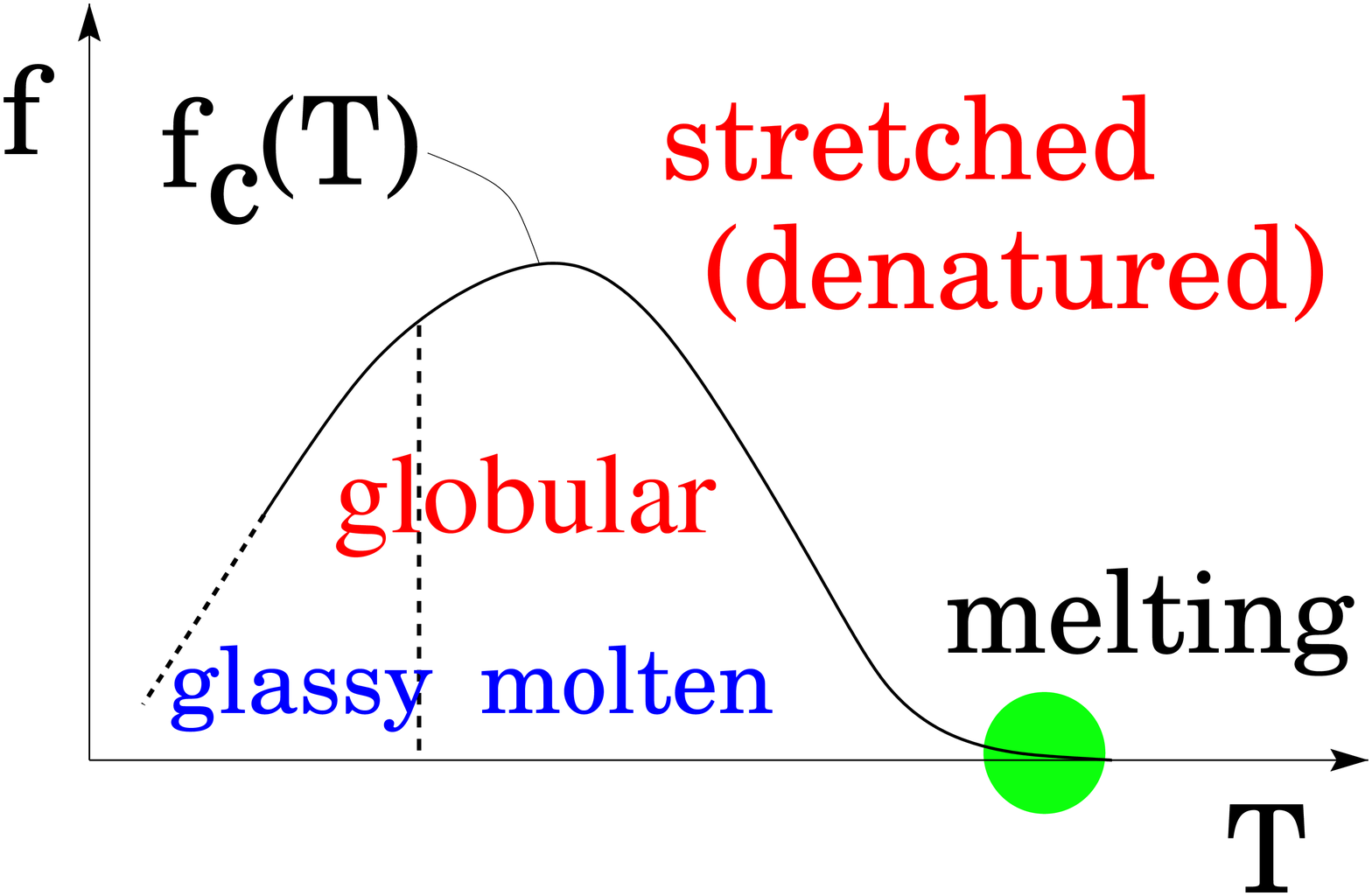}}
\caption{Phase diagram of disordered RNA as a function of temperature and the force. The molecule undergoes a continuous opening transition at a critical force $f_c(T)$ that we predict to be non-monotonous as a function of temperature. This gives rise to a re-entrance phenomenon at fixed forces in a certain interval. At low temperature the system is in a glassy phase characterized by a small number of low-lying metastable states. At higher temperatures, RNA is in a molten state, that behaves in essentially the same way as a homopolymer. There is a thermal denaturation transition if the entropic penalties for loops are sufficiently large. Otherwise there is simply a crossover, and $f_c(T)$ never really vanishes.}
\label{Figphasedia}
\end{figure}

The different behavior at low and high temperatures implies that the critical force reaches a maximum somewhere below the denaturation temperature which gives rise to the following re-entrance phenomenon (see Fig.~\ref{Figphasedia}): If one fixes the external force at a value smaller than the maximum of $f_c(T)$ and then decreases the temperature, starting in the denatured regime, the molecule will collapse to the globular state when it crosses the critical line for the first time. However, it will re-enter the stretched phase again at lower temperature when the second crossing occurs. Since the transitions are of second order this behavior would be seen as a (continuous) breathing of the molecule upon changing the temperature. We expect this effect to be relevant for disordered RNA-sequences where $t=0$ applies as we argued above. Such a behavior has been seen in numerical simulations of protein unfolding \cite{KlimovThirumalai01}. A similar effect was also predicted in the form of 'cold denaturation' in DNA unzipping \cite{Orlandini01, Marenduzzo02c}. 

\subsection{The opening crossover above $f_c$}

\begin{figure}[!]

\resizebox{0.5\textwidth}{!}{\includegraphics{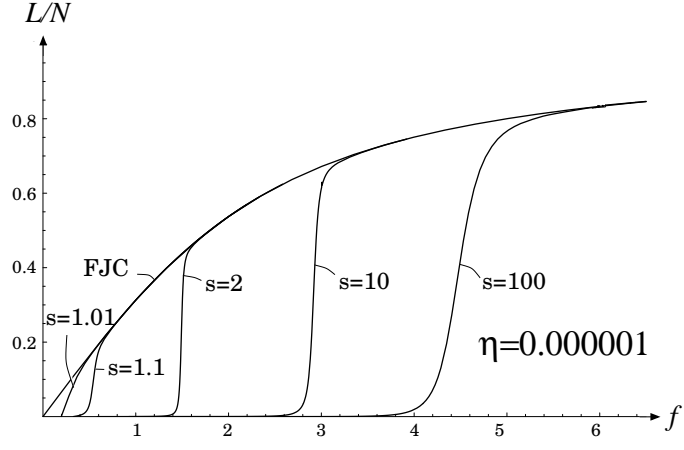}}
\caption{$f-l$-curves at large cooperativity: The initial linear slope at the transition is proportional to $\eta$. The phase transition is masked by the subsequent crossover whose steepness scales as $\eta^{-1/2}$, independently of $s$.}
\label{eta0.000001}
\end{figure}

\begin{figure}[!]

\resizebox{0.5\textwidth}{!}{\includegraphics{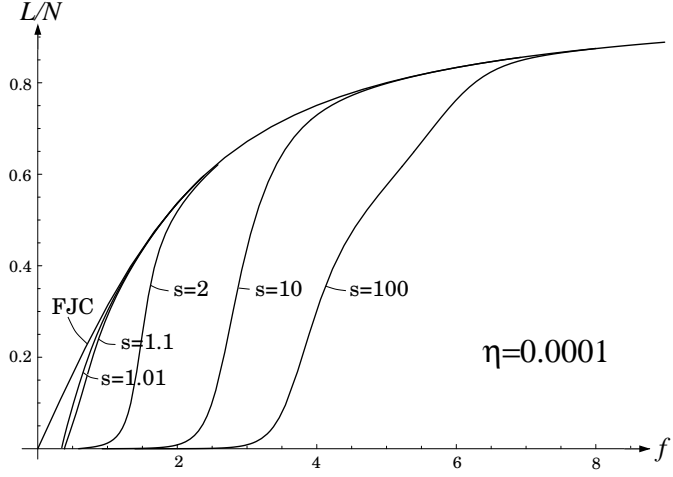}}
\caption{$f-l$-curves at intermediate cooperativity: Critical point and crossover merge for sufficiently small values of $s$. }
\label{eta0.0001}
\end{figure}

\begin{figure}[!]

\resizebox{0.5\textwidth}{!}{\includegraphics{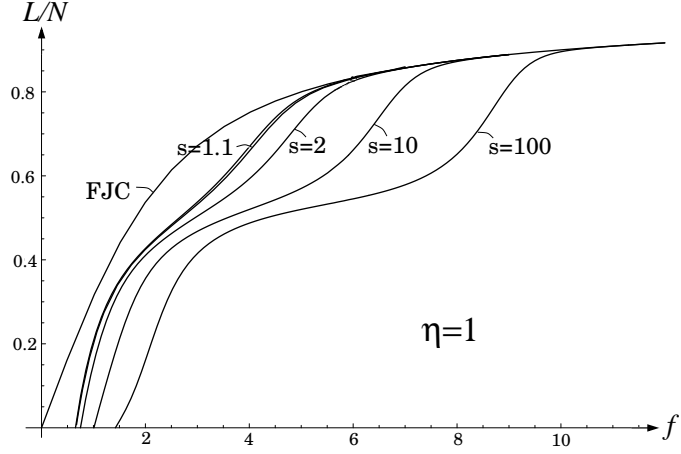}}
\caption{$f-l$-curves with no cooperativity: First the large closed structures open up to form smaller units. The plateau corresponds to necklaces of hairpins of the smallest possible size. Those structures are disrupted only at a higher force. This behavior is an artifact of very strong pairing energies.}
\label{eta1}
\end{figure}

In the following, we discuss several features of the force-extension curves in the thermodynamic limit, in particular we will derive how the linear slope above the critical point and the characteristics of the subsequent crossover depend on the energy parameters of the model. The results are illustrated in Figs.~\ref{eta0.000001},\ref{eta0.0001} and \ref{eta1} where we plot force-extension curves for various pairs of $s$ and $\eta$. The structural parameters are those appropriate for homogeneous RNA ($n=3$, $t=3$, and $l_b=4 l$).
 
As we discussed in section \ref{force}, the extension $L(f)$ of the molecule slightly above the critical force grows like $N(f-f_c)$. The prefactor and the range of validity of the linear regime can be calculated from an expansion of Eq.~(\ref{zetaf2}) around the critical point. In the different temperature regimes, we find
\be
	\label{lincoeff}
	L(f)/(N(f-f_c))\sim
	\left\{ \begin{array}[c]{ll}
	O\left(\eta s^{-t/4}\right) & s\gg 1,   \\
	O\left(\eta/(s-1)^{4}\right) & 1\gg s-1\gg \eta^{1/3},\\
	O(\eta^{-1/3}) & s-1\ll\eta^{1/3}.
	\end{array}	\right.  
\ee
valid within a force window 
\be
	\label{linregf}
	f-f_c=\left\{ \begin{array}[c]{ll}
	O\left(1\right) & s\gg 1, \\
	O\left(s-1\right) & 1\gg s-1\gg \eta^{1/3},\\
	O(\eta^{1/3}) & s-1\ll\eta^{1/3}.
	\end{array}	\right. 
\ee



Note that at low temperatures ($s\gg 1$) the extension is suppressed by a 
factor $\eta$ (see Fig.~\ref{eta0.000001}). This is a consequence of the cooperativity in the system, 
i.e., the tendency to form long helices and large structures. Their size 
increases typically as $\eta^{-1}$, so that necklaces composed of such 
structures are very short. The effect is even enhanced for $t>0$ where it 
is favorable to form longer helices in order to reduce the energy loss 
in terminal loops. The extension grows linearly with the force over a range of
order $O(1)$ which is comparable to the critical force itself. 
This is in contrast to the case near denaturation where the extent of the 
linear regime is much smaller than $f_c$ and the coefficient of the 
linear term can be appreciably large.

At higher forces the force-extension characteristics becomes non-linear, 
and finally, the majority of base pairings opens up in a sharp crossover. 
This happens at a force $f_*$ where the free energy of bases in the free 
part of the chain equals half of the base pairing free energy, 
$\zeta_{\rm ss}(f_*)\approx s^{-1/2}$. Note that this is the analog of the critical force in DNA unzipping which is of a different nature than $f_c$. Beyond $f_*$, the molecule behaves essentially  
as a freely jointed chain.

The crossover is most interesting at low temperatures, $s \gg 1$. In the appendix it is shown to take place within a force window determined by $\rho(f)\equiv[\eta s^{-(1+t)/2}/\zeta_b(f)]^{1/2}/|\zeta_{\rm ss}(f)-s^{-1/2}| = O(1)$. Its width scales like 
\be
	\label{forcewindow}
	\Delta f\sim \left(\eta s^{1+(l_b/l-t-1)/2}\right)^{1/2}\approx  \left(\eta s\right)^{1/2}.
\ee
where we assumed $l_b/l\approx t+1$ for homogeneous RNA. This expresses the fact that the smallest possible hairpin loop (with length $l\cdot(t+1)$) will almost be a direct, stretched bridge between the bases of the adjacent pair (at distance $l_b$). The crossover is very sharp if the cooperativity, as measured by $\eta^{-1}$, is high.

Below the crossover ($\rho(f) \ll 1$) we find the extension to be small as compared to that of a freely jointed chain subject to the same force,
\be 
	\label{lbelowcross}
	\frac{L(f)}{N}\approx \frac{L_{\rm FJC}(f)}{N} \rho(f).
\ee
Above the crossover ($\rho(f) \gg 1$), the role played by closed structures is negligible, and the force-extension curve approaches that of a freely jointed chain,
\be 
	\label{labovecross}
	\frac{L(f)}{N}\approx \frac{L_{\rm FJC}(f)}{N} \left[1+O\left(\frac{1}{\rho(f)} \right)\right].
\ee

Near denaturation ($s\approx 1$), the situation is less interesting since both the phase transition and the final opening of the secondary structures take place at very small forces. Still, there is a discernible crossover in the case $s-1\gg \eta^{1/3}$ that occurs at a force $f_*\approx f_c+O(s-1)$ while its width scales like $\Delta f\sim [\eta/(s-1)]^{1/2}$. In the limit $s-1\ll\eta^{1/3}$, however, no crossover can be seen since all characteristic force scales behave as $\eta^{1/3}$.

\section{Conclusion}
We have studied several aspects of RNA folding on the level of the secondary structure. We concentrated on a homogeneous model that we expect to describe random sequences on a coarse-grained level as well, provided that the parameters are appropriately renormalized. This point of view is strongly supported by the finding that, at sufficiently high temperatures, disorder is an irrelevant perturbation for the force-induced opening transition in the sense that scaling functions for disordered sequences superpose perfectly with the analytical curve calculated in the homogeneous case. 

Our model takes into account base pairing and stacking energies, as well as entropic costs for loops. The latter have been shown to give rise to a thermodynamic denaturation transition if they are sufficiently strong. However, even when loop penalties are included, the dominant secondary structures in the collapsed phase of large molecules are too dense to be accommodated in three-dimensional space. Even at moderate, biologically relevant sizes, it is questionable whether the usual neglect of excluded volume (and other tertiary) interactions in the prediction of secondary structure is justified. 

The force-induced unfolding has been studied in detail. We have characterized the second order phase transition separating a globular from a necklace-like extensive phase. A correlation length diverging like $(f-f_c)^{-2}$ has been identified as the typical size of the largest closed structures that appear in the necklace above $f_c$. The critical scaling of the correlation length remains (marginally) unchanged upon introduction of disorder, as follows from a Harris-type criterion. However, in the low temperature, glassy phase the other critical exponents are modified, in contrast to the high temperature phase which belongs to the same universality class as a homogeneous polymer. This difference of the temperature regimes manifests itself as well in the force-extension characteristics. In the present paper, we have restricted ourselves to the discussion of the homogeneous or high temperature case. The jump-like force-extension curves in the glassy regime are discussed in \cite{MullerKrzakalaMezard02}.

The second order phase transition in force-induced unfolding has been shown to be a very robust feature of the model. In particular, it is independent of the specific pairing and stacking energies and further structural parameters. It occurs at a critical force that we predict to be a non-monotonic function of temperature in the case of disordered sequences. This gives rise to a re-entrance phenomenon when the temperature is varied at constant force.

When the stacking energy is large, the pairing behavior is highly cooperative. This can render the second order phase transition almost invisible, since the extension grows very slowly as a function of $f-f_c$. On the other hand, a very sharp first-order-like crossover occurs at the (higher) force where essentially all base pairs are disrupted.

\acknowledgments

It is a pleasure to thank Florent Krzakala, Marc M{\'e}zard and Andrea Montanari for many helpful discussions. I am indebted to Marc M{\'e}zard for carefully reading the manuscript. 

I acknowledge a fellowship from the MRT. The LPTMS is an Unit\'e de Recherche de
l'Universit\'e Paris~XI associ\'ee au CNRS.

\section*{Appendix}
\label{appendix}
In this appendix we outline in more detail how to obtain the characteristics of the force-extension curves of section \ref{extpart}. 

\subsection*{The singularities of the partition function}

Let us start from Eq.~(\ref{zetaf2})
\be
	\label{zetaf2app}
	\zeta_{\rm ss}(f)=\zeta(f)+\frac{\Xi_c(\zeta(f))}{\zeta_b(f)},
\ee
which determines the pole singularity $\zeta(f)$ as a function of the force for $f>f_c$. The functional form of the generating function $\Xi_c(\zeta)$ for closed structures can be obtained from Eq.~(\ref{withoutloop}) as
\be 
	\label{genFwl}
	 \Xi_c(\zeta)=\frac{1}{2}\left(1-\zeta-\frac{\eta(s\zeta^2)^n(1-\zeta^t)/(1-\zeta)}{1-s\zeta^2+\eta(s\zeta^2)^n}\right)-\sqrt{\frac{1}{4}\left(1-\zeta-\frac{\eta(s\zeta^2)^n(1-\zeta^t)/(1-\zeta)}{1-s\zeta^2+\eta(s\zeta^2)^n}\right)^2-\frac{\eta(s\zeta^2)^n\zeta^t}{1-s\zeta^2+\eta(s\zeta^2)^n}}.
\ee

The free energy per base of the globular phase is related to the singularity $\zeta_*$ of $\Xi_c$ via $\zeta_*=\exp[-\beta\phi_c]$. $\zeta_*$ is given by the vanishing of the square root in Eq.~(\ref{genFwl}). In the low temperature regime, $s\gg1$, we find

\be
	\label{zetaclowT}
  \zeta_*\approx\left\{ \begin{array}[c]{ll}
	s^{-1/2}\left(1-\frac{3\eta}{2}\right) & t=0, \\
	s^{-1/2}\left(1-\frac{\eta}{s^{1/4}}\right) & t=1, \\
	s^{-1/2}\left(1-\frac{2\eta}{s^{1/2}}\right) & t=2, \\
	s^{-1/2}\left(1-\frac{\eta}{s^{1/2}}\right) & t\ge3, \\
	\end{array}	\right.
\ee

while near the denaturation, $s-1\ll1$, we obtain

\be
	\label{zetachighT}
	\zeta_*\approx\left\{ \begin{array}[c]{ll}
	s^{-1/2}\left(1-\frac{8\eta}{(s-1)^2}\right) & 1\gg s-1\gg \eta^{1/3},\\
	1-(2\eta)^{1/3} & s-1\ll\eta^{1/3}.
	\end{array}	\right. 
\ee
We recall that we assume $\eta \ll 1$ throughout.

\subsection*{The critical force}

The critical force $f_c$ has to be determined from the crossing of the two singularities, $\zeta(f_c)=\zeta_*$, or more explicitly,
\be
\label{fccond}
\zeta_{\rm ss}(f_c)=\zeta_*+\Xi_c(\zeta_*)/\zeta_b(f_c),
\ee
where we can use the fact that the square root in (\ref{genFwl}) vanishes at $\zeta_*$,
\be
	\nonumber
	\Xi_c(\zeta_*)=\sqrt{ \frac{\eta(s\zeta^2)^n\zeta^t}{1-s\zeta^2+\eta(s\zeta^2)^n} }.
\ee

In the low temperature regime, $s\gg 1$, we have to distinguish different values of $t$. In the case $t\ge 1$ we find approximately 

\be 
	\label{fclowT}
	\zeta_{\rm ss}(f_c(T)) \approx \zeta_*+\frac{\zeta_*^{t/2}}{\zeta_b(f_c(T))}.
\ee

We can neglect the first term on the right hand side for $t<2(l_b/l+1)$ which is always satisfied for homogeneous RNA where we have $t\approx l_b/l-1$.

Recalling the definition $s=\exp(\beta| f_{\rm pair}(T)|)$ and approximating $\zeta_{\rm ss}(f)\equiv \beta l f/\sinh( \beta l f)\approx \exp(-\beta l f)$ and $\zeta_{b}(f)\equiv \beta l_b f/\sinh( \beta l_b f)\approx \exp(-\beta l_b f)$ at low temperatures, we find 
\be
	\label{fcexplicitapp}
	f_c(T)\approx\frac{t}{4(l_b+l)}|f_{\rm pair}(T)|.
\ee 	

The behavior for $t=0$ is different since all bases can be paired, even those at the end of a hairpin. In that case, (\ref{fccond}) reduces to 
\be
	\label{fct=0app}
	2\zeta_{\rm ss}(f_c)\zeta_{b}(f_c)\approx 1,
\ee
implying $f_c(T)\propto T/l$, almost independently of $s$. This underlines the purely entropic origin of the critical force sufficiently below denaturation .

In the regime $s\approx 1$ the critical force is small, decreasing as

\be
	\label{fchighTapp}
	f_c(T)=\left\{ \begin{array}[c]{ll}
	 O((s-1)^{1/2}) & 1\gg s-1\gg \eta^{1/3}\\
	 O(\eta^{1/6}) & s-1\ll\eta
	\end{array}	\right. 
\ee
on approaching the denaturation. This follows simply from an expansion of (\ref{fccond}) at low forces.

\subsection*{The linear regime above $f_c$}

In the thermodynamic limit the force-extension curve starts off linearly from zero extension at $f_c$. To obtain the slope of the curve, we expand Eq.~(\ref{zetaf2app}) around the critical point. In particular, we have to expand the square root in Eq.~(\ref{genFwl}) at $\zeta_*$,
\be
	\label{znearzc}
	\Xi_c(\zeta)\approx 
	\frac{1}{2}\left(1-\zeta-\frac{\eta(s\zeta^2)^n}{1-s\zeta^2+\eta(s\zeta^2)^n(1-\zeta^t)/(1-\zeta)}\right)
	-\sqrt{A(\zeta_*-\zeta)+B(\zeta_*-\zeta)^2}.
\ee

As long as we can neglect $B(\zeta_*-\zeta)^2$ with respect to $A(\zeta_*-\zeta)$, i.e., for $\zeta_*-\zeta(f)\ll A/B$, we may obtain $\zeta(f)$ approximately from (see Eq.~(\ref{zetaf2app}))
\be 
	\label{zetaflinear}
	\zeta_{\rm ss}(f)\zeta_b(f)|_{f_c}^f\approx (f-f_c)[\zeta_{\rm ss}\zeta_b]^\prime(f_c)
	\stackrel{!}{\approx} 
	\Xi_c(\zeta)|_{\zeta_*}^{\zeta(f)}\approx -[A(\zeta_*-\zeta(f))]^{1/2},
\ee
and thus,
\be 
	\label{zetaflinear2}
	\zeta(f)\approx \zeta_*-\frac{(f-f_c)^2}{A}[\zeta_{\rm ss}\zeta_b]^{\prime2}(f_c).
\ee
The extension in the linear scaling regime above $f_c$ then follows via
$L(f)=-N\beta^{-1}\partial\ln[\zeta(f)]/\partial f\sim N(f-f_c)$. In the different temperature regimes, the coefficient of $N(f-f_c)$ scales like
\be
	\label{lincoeffapp}
	L(f)/(N(f-f_c))\sim
	\left\{ \begin{array}[c]{ll}
	O\left(\eta s^{-t/4}\right) & s\gg 1,  \\
	O\left(\eta/(s-1)^{4}\right) & 1\gg s-1\gg \eta^{1/3},\\
	O(\eta^{-1/3}) & s-1\ll\eta^{1/3}.
	\end{array}	\right.  
\ee
The linear regime extends up to forces determined by $\zeta_*-\zeta(f)\approx A/B$, from which we obtain the force windows 
\be
	\label{linregfapp}
	f-f_c=\left\{ \begin{array}[c]{ll}
	O\left(1\right) & s\gg 1, \\
	O\left(s-1\right) & 1\gg s-1\gg \eta^{1/3},\\
	O(\eta^{1/3}) & s-1\ll\eta^{1/3}.
	\end{array}	\right. 
\ee

\subsection*{The non-linear regime and the crossover}

At higher forces the force-extension curve becomes non-linear, but the typical size of closed structures is still large so that the second term on the right-hand side of (\ref{zetaf2app}) cannot be neglected. Only in a later stage the base pairs open up completely, and the molecule becomes a freely jointed chain.

In the further discussion of the characteristics of the force-extension curve we will restrict ourselves to (real) homogeneous RNA with $1<t\approx l_b/l-1$. We will only treat the case $s\gg1$ in some detail, the case $s\approx 1$ can be treated analogously.

Beyond the linear regime, we can solve for $\zeta(f)$ using the following approximation: The generating function for closed structures can be simplified by expanding the square root in Eq.~(\ref{genFwl}) with respect to the second term
\be
	\label{squarerootexp}
	\Xi_c(\zeta(f))\approx 
	\frac{\eta(s\zeta^2)^n\zeta^t}{(1-\zeta)[1-s\zeta^2+\eta(s\zeta^2)^n]-\eta(s\zeta^2)^n\frac{1-\zeta^t}{1-\zeta}}.
\ee
The (\ref{zetaf2}) for $\zeta(f)$ then reduces to a quadratic equation up to terms of the order of $\sqrt{\eta}$, except in a narrow region around the force $f_*$ where the opening crossover takes places. The crossover force $f_*$ is approximately determined by $\zeta_{\rm ss}(f_*)=s^{-1/2}$.

Before proceeding let us estimate the width of the crossover. Considering the right hand side of Eq.~(\ref{zetaf2app}), we note that the closed structures are important as long as the variation of $\Xi_c(\zeta(f))/\zeta_b(f)$ with force dominates that of $\zeta(f)$. The crossover to the freely jointed chain regime occurs when both variations become comparable. Let us therefore write $\zeta(f)=s^{-1/2}(1-\epsilon(f))$ and determine the value $\epsilon_X$ at which the correction $s^{-1/2}\epsilon_X$ begins to dominate the term $\Xi_c(\zeta(f))/\zeta_b(f)$. Using Eq.~(\ref{squarerootexp}) and the fact that $\eta/\epsilon_X$ will be small, we find $\Xi_c\approx s^{-t/2}\eta/\epsilon_X$. Equating this to $s^{-1/2}\epsilon_X\zeta_b(f)$, we find 
\be
	\label{crosswidth}
	s^{-1/2}\epsilon_X\equiv s^{-1/2}-\zeta(f)\sim [\eta s^{-(1+t)/2}/\zeta_b(f)]^{1/2}. 
\ee

More quantitatively, one finds that for forces such that $\zeta_{\rm ss}(f)-s^{-1/2}\gg[\eta s^{-(1+t)/2}/\zeta_b(f)]^{1/2}$, $\zeta(f)$ is given by
\be 
	\label{zetabelowcross}
	\zeta(f)\approx s^{-1/2}\left(1-\frac{\eta s^{-t/2}}{2\zeta_b(f)(\zeta_{\rm ss}(f)-s^{-1/2})}\right).
\ee
The extension follows from a logarithmic derivative,
\be 
	\label{lbelowcrossapp}
	\frac{L(f)}{N}\approx \frac{L_{\rm FJC}}{N} \frac{\eta s^{-t/2}\zeta_{\rm ss}(f)}{\zeta_b(f)(\zeta_{\rm ss}(f)-s^{-1/2})^2},
\ee
which is small as compared to that of a freely jointed chain at the same force.

For forces such that $\zeta_{\rm ss}(f)-s^{-1/2}\ll-[\eta s^{-(1+t)/2}/\zeta_b(f)]^{1/2}$ closed structures play a negligible role. To the same degree of approximation as before one finds
\be 
	\label{zetaabovecross}
	\zeta(f)\approx \zeta_{\rm ss}(f) \left[1+O\left(\frac{\eta s^{-(1+t)/2}}{\zeta_b(f)(\zeta_{\rm ss}(f)-s^{-1/2})}\right)\right],
\ee
and the force-extension curve joins that of a freely jointed chain
\be 
	\label{labovecrossapp}
	\frac{L(f)}{N}\approx \frac{L_{\rm FJC}(f)}{N} \left[1+O\left(\frac{\eta s^{-(1+t)/2}}{\zeta_b(f)(\zeta_{\rm ss}(f)-s^{-1/2})}\right)\right].
\ee

The force window over which the crossover takes place derives from (\ref{crosswidth}) and scales like
\be
	\label{forcewindowapp}
	\Delta f\sim \left(\eta s^{1+(l_b/l-t-1)/2}\right)^{1/2}.
\ee

For the case near denaturation, $s\approx 1$ the calculations are analogous and yield a crossover around $f_*\approx f_c+O(s-1)$ with a width scaling like $\Delta f\sim [\eta/(s-1)]^{1/2}$ for $s-1\gg \eta^{1/3}$. In the limit $s-1\ll\eta^{1/3}$, however, all characteristic force scales behave as $\eta^{1/3}$, and a separation of into different regimes does not make sense.

\bibliographystyle{prsty}
\bibliography{../Bib/references}

\addcontentsline{toc}{chapter}{\protect\bibname}
\begin{thebibliography}{10}

\bibitem{DeGennes68}
P.-G. de~Gennes, Bioploymers {\bf 6},  715  (1968).

\bibitem{BustamanteSmith96}
S.~B. Smith, Y. Cui, and C. Bustamante, Science {\bf 271},  795  (1996).

\bibitem{Maier00}
B. Maier, D. Bensimon, and V. Croquette, Proc. Natl. Acad. Sci. USA {\bf 97},
  12002  (2000).

\bibitem{BockelmannThomen02}
U. Bockelmann {\it et~al.}, Biophys J. {\bf 82},  1537  (2002).

\bibitem{MontanariMezard01}
A. Montanari and M. M{\'e}zard, Phys. Rev. Lett. {\bf 86},  2178  (2001).

\bibitem{ZhouZhang01}
H. Zhou and Y. Zhang, J. Chem. Phys. {\bf 114},  8694  (2001).

\bibitem{BustamanteTinoco99}
I. Tinoco and C. Bustamante, J. Mol. Biol. {\bf 293},  271  (1999).

\bibitem{ZukerSankoff84}
M. Zuker and D. Sankoff, Bull. Math. Biol. {\bf 46},  591  (1984).

\bibitem{Schuster94}
I.~L. Hofacker {\it et~al.}, Monatsh. Chem. {\bf 125},  167  (1994).

\bibitem{Tinoco71}
I. Tinoco, O.~C. Uhlenbeck, and M.~D. Levine, Nature {\bf 230},  362  (1971).

\bibitem{Tinoco73}
I. Tinoco {\it et~al.}, Nature New Biology {\bf 246},  40  (1973).

\bibitem{Zukermfold}
D.~H. Mathews, J. Sabina, M. Zuker, and D.~H. Turner, J. Mol. Biol. {\bf 288},
  911  (1999).

\bibitem{Duplantier86}
B. Duplantier, Phys. Rev. Lett. {\bf 57},  941  (1986).

\bibitem{Schaefer92}
L. Sch{\"a}fer, C. von Ferber, U. Lehr, and B. Duplantier, Nucl. Phys. B {\bf
  374},  473  (1992).

\bibitem{KafriMukamel00}
Y. Kafri, D. Mukamel, and L. Peliti, Phys. Rev. Lett. {\bf 85},  4988  (2000).

\bibitem{KafriMukamel01a}
Y. Kafri, D. Mukamel, and L. Peliti, Eur. Phys. J. B {\bf 27},  135  (2002).

\bibitem{MetzlerHanke02}
R. Metzler {\it et~al.}, Phys. Rev. E {\bf 65},  061103  (2002).

\bibitem{BundschuhHwa01b}
R. Bundschuh and T. Hwa, Phys. Rev. E {\bf 65},  031903  (2002).

\bibitem{Nienhuis82}
B. Nienhuis, Phys. Rev. Lett. {\bf 49},  1062  (1982).

\bibitem{Coniglio89}
A. Coniglio, Phys. Rev. Lett. {\bf 62},  3054  (1989).

\bibitem{Woodson00}
S.~A. Woodson, Nat. Struct. Biol. {\bf 7},  349  (2000).

\bibitem{Russell00}
R. Russell, I.~S. Millett, S. Doniach, and D. Herschlag, Nat. Struct. Biol.
  {\bf 7},  367  (2000).

\bibitem{ThirumalaiWoodson96}
D. Thirumalai and S.~A. Woodson, Acc. Chem. Res. {\bf 29},  433  (1996).

\bibitem{Higgs96}
P.~G. Higgs, Phys. Rev. Lett. {\bf 76},  704  (1996).

\bibitem{BundschuhHwa01a}
R. Bundschuh and T. Hwa, "Europhys. Lett." {\bf 59},  903  (2002).

\bibitem{PagnaniParisi00}
A. Pagnani, G. Parisi, and F. Ricci-Tersenghi, Phys. Rev. Lett. {\bf 84},  2026
   (2000).

\bibitem{PagnaniParisi00Reply}
A. Pagnani, G. Parisi, and F. Ricci-Tersenghi, Phys. Rev. Lett. {\bf 86},  1383
   (2001).

\bibitem{Hartmann01}
A. Hartmann, Phys. Rev. Lett. {\bf 86},  1382  (2001).

\bibitem{KrzakalaMezardMuller02}
F. Krzakala, M. M{\'e}zard, and M. M{\"u}ller, Europhys. Lett. {\bf 57},  752
  (2002).

\bibitem{IsambertSiggia00}
H. Isambert and E.~D. Siggia, Proc. Natl. Acad. Sci. USA {\bf 97},  6515
  (2000).

\bibitem{LubenskyNelson00}
D.~K. Lubensky and D.~R. Nelson, Phys. Rev. Lett. {\bf 85},  1572  (2000).

\bibitem{LubenskyNelson02}
D.~K. Lubensky and D.~R. Nelson, Phys. Rev. E {\bf 65},  031917  (2002).

\bibitem{GerlandBundschuh01}
U. Gerland, R. Bundschuh, and T. Hwa, Biophys. J. {\bf 81},  1324  (2001).

\bibitem{MullerKrzakalaMezard02}
M. M{\"u}ller, F. Krzakala, and M. M{\'e}zard, Eur. Phys J. E {\bf 9},  67
  (2002).

\bibitem{MarinariPagnaniRicci02}
E. Marinari, A. Pagnani, and F. Ricci-Tersenghi, Phys. Rev. E {\bf 65},  041919
   (2002).

\bibitem{Waterman78}
M. Waterman, Adv. Math. Suppl. Studies {\bf 1},  167  (1978).

\bibitem{McCaskill90}
J.~S. McCaskill, Biopolymers {\bf 29},  1105  (1990).

\bibitem{GeisslerShakhnovich02a}
P.~L. Geissler and E.~I. Shakhnovich, Phys. Rev. E {\bf 65},  056110  (2002).

\bibitem{GeisslerShakhnovich02b}
P.~L. Geissler and E.~I. Shakhnovich, Macromolecules {\bf 35},  4429  (2002).

\bibitem{Marenduzzo02b}
F.~S. D.~Marenduzzo, A.~Maritan, J. Phys. A {\bf 35},  L233  (2002).

\bibitem{KlimovThirumalai01}
D.~K. Klimov and D. Thirumalai, J. Phys. Chem. B {\bf 105},  6648  (2001).

\bibitem{Orlandini01}
E. Orlandini {\it et~al.}, J. Phys. A {\bf 34},  L751  (2001).

\bibitem{Marenduzzo02c}
D. Marenduzzo {\it et~al.}, Phys. Rev. Lett. {\bf 88},  028102  (2002).

\end{thebibliography}

\end{document}